%% file: _main.tex
\input{_constants}
\arxiv
\pdfoutput=1
\documentclass[10pt,twocolumn,letterpaper]{article}
\input{cvpr_header}

\title{\paperTitle}

\author{\authorBlock} 
\begin{document}

\twocolumn[{%
            \renewcommand\twocolumn[1][]{#1}%
            \maketitle
            \input{figs/teaser}

        }]

\ifcamera {\renewcommand{\thefootnote}{\fnsymbol{footnote}} \footnotetext[2]{Work done during internship at Stability AI.}} \fi

\ifarxiv {\renewcommand{\thefootnote}{\fnsymbol{footnote}} \footnotetext[2]{Work done during internship at Stability AI.}} \fi

\input{00_abstract}

\input{01_intro}

\input{02_related}

\input{03_method}

\input{04_results}

\input{10_conclusion}

{\small
\bibliographystyle{ieeenat_fullname}
\bibliography{11_references}
}

\ifarxiv \clearpage \appendix \input{12_appendix} \fi

\end{document}

%% file: _constants.tex
\def\paperID{631}
\def\confName{ICCV}
\def\confYear{2025}

\def\shortTitle{SViM3D}
\def\longerTitle{Stable Video Materials 3D}

\def\paperTitle{SViM3D: Stable Video Material Diffusion for Single Image  3D Generation}

\def\authorBlock{
Andreas Engelhardt$^{1,2{\footnotemark[2]}}$
\quad
Mark Boss$^1$
\quad
Vikram Voleti$^1$
\quad
Chun-Han Yao$^1$
\quad
\\
Hendrik P. A. Lensch$^2$
\quad
Varun Jampani$^1$ \\
\\
{\small $^1$Stability AI \quad $^2$University of T{\"{u}}bingen}
}

\newif\ifreview 
\newif\ifarxiv \newcommand{\arxiv}{\arxivtrue}
\newif\ifcamera 
\newif\ifrebuttal 

\newif\ifdrafting
\draftingtrue %
\ifdrafting
\newcommand{\aen}[1]{{\color{blueish} [AE: #1]}} %
\newcommand{\mb}[1]{{\color{blueish}#1}} %
\newcommand{\MB}[1]{{\color{blueish}{\bf [MB: #1]}}} %
\newcommand{\Mb}[1]{\marginpar{\tiny{\textcolor{blueish}{#1}}}} %
\newcommand{\vj}[1]{{\color{dark_green}{\bf [VJ: #1]}}}
\newcommand{\VJ}[1]{{\color{dark_green}{\bf [VJ: #1]}}}
\newcommand{\Vj}[1]{\marginpar{\tiny{\textcolor{dark_green}{#1}}}}

\newcommand{\hl}[1]{{\color{brown}{\bf [HL: #1]}}}

\newcommand{\VV}[1]{\textcolor{teal}{\textbf{[VV: #1]}}}

\newcommand{\ds}[1]{{\color{red}[DS: #1]}}
\newcommand{\todo}[1]{{\color{red}{\textbf{TODO}: #1}}}
\else
\newcommand{\aen}[1]{}
\newcommand{\mb}[1]{} 
\newcommand{\MB}[1]{} 
\newcommand{\Mb}[1]{} 
\newcommand{\vj}[1]{}
\newcommand{\VJ}[1]{}
\newcommand{\Vj}[1]{}
\newcommand{\jb}[1]{}
\newcommand{\JB}[1]{}
\newcommand{\Jb}[1]{}
\newcommand{\ar}[1]{}
\newcommand{\AR}[1]{}
\newcommand{\Ar}[1]{}
\newcommand{\todo}[1]{}
\newcommand{\TODO}[1]{}
\newcommand{\hl}[1]{}
\newcommand{\ds}[1]{}
\newcommand{\VV}[1]{}
\fi

%% file: cvpr_header.tex
\ifarxiv \usepackage[pagenumbers]{cvpr} \fi
\ifrebuttal \usepackage[rebuttal]{cvpr} \fi
\ifcamera \usepackage{cvpr} \fi

\input{_macros}  %

\usepackage{xr-hyper}

\makeatletter
\newcommand*{\addFileDependency}[1]{
  \typeout{(#1)}
  \@addtofilelist{#1}
  \IfFileExists{#1}{}{\typeout{No file #1.}}
}

\makeatother

\definecolor{cvprblue}{rgb}{0.21,0.49,0.74}
\usepackage[pagebackref,breaklinks,colorlinks,citecolor=cvprblue]{hyperref}
\usepackage[capitalize]{cleveref}
\crefname{section}{Sec.}{Secs.}
\crefname{table}{Table}{Tables}
\crefname{figure}{Fig.}{Figs.}

%% file: _macros.tex
\usepackage{graphicx}	
\usepackage{amsmath}	
\usepackage{amssymb}	
\usepackage{booktabs}
\usepackage{colortbl}
\usepackage{times}
\usepackage{microtype}
\usepackage{epsfig}
\usepackage[table,xcdraw,dvipsnames]{xcolor}
\usepackage{caption}
\usepackage{float}
\usepackage{placeins}
\usepackage{color, colortbl}
\usepackage{stfloats}
\usepackage{enumitem}
\usepackage{tabularx}
\usepackage{xstring}
\usepackage{multirow}
\usepackage{xspace}
\usepackage{xfrac}
\usepackage{url}
\usepackage{subcaption}
\usepackage{colortbl}
\usepackage{bm}
\usepackage[hang,flushmargin]{footmisc}
\usepackage{lineno}
\usepackage{siunitx}

\ifcamera \usepackage[accsupp]{axessibility} \fi

\usepackage{tikz}

\usetikzlibrary{shapes}
\usetikzlibrary{calc}
\usetikzlibrary{backgrounds}
\usetikzlibrary{positioning}
\usetikzlibrary{arrows}
\usetikzlibrary{arrows.meta}
\usetikzlibrary{decorations.text}    
\usetikzlibrary{fit}
\usetikzlibrary{spy}
\usetikzlibrary{3d}
\usetikzlibrary{math}

\definecolor{turquoise}{cmyk}{0.65,0,0.1,0.3}
\definecolor{purple}{rgb}{0.65,0,0.65}
\definecolor{dark_green}{rgb}{0, 0.5, 0}
\definecolor{orange}{rgb}{0.8, 0.6, 0.2}
\definecolor{red}{rgb}{0.8, 0.2, 0.2}
\definecolor{darkred}{rgb}{0.6, 0.1, 0.05}
\definecolor{blueish}{rgb}{0.0, 0.3, .6}
\definecolor{light_gray}{rgb}{0.7, 0.7, .7}
\definecolor{pink}{rgb}{1, 0, 1}
\definecolor{greyblue}{rgb}{0.25, 0.25, 1}

\definecolor{bestcol}{RGB}{254,196,79}
\newcommand{\best}[1]{\cellcolor{bestcol} \textbf{#1}}
\definecolor{secondbestcol}{RGB}{255,247,188}
\newcommand{\secondbest}[1]{\cellcolor{secondbestcol} #1}

\newcommand{\OURS}{\shortTitle~}

\ifarxiv  \fi

\newcommand{\R}[1]{{%
    \textbf{%
        \ifstrequal{#1}{1}{\textcolor{red}{R1}}{%
        \ifstrequal{#1}{2}{\textcolor{blueish}{R2}}{%
        \ifstrequal{#1}{3}{\textcolor{purple}{R3}}{%
        \ifstrequal{#1}{4}{\textcolor{teal}{R#1}}{%
                           \textcolor{cyan}{R#1}%
        }}}}%
    }%
}}

\newcommand{\vect}[1]{\bm{#1}}

\newcommand\rgt{\aftergroup\mathclose\aftergroup{\aftergroup}\right}

\newcommand{\fig}[1]{Fig.~\ref{fig:#1}}

\newcommand{\eqn}[1]{Eq.~\ref{eq:#1}}

\makeatletter
\DeclareRobustCommand\onedot{\futurelet\@let@token\@onedot}
\def\@onedot{\ifx\@let@token.\else.\null\fi\xspace}

\def\ie{\emph{i.e}\onedot}

\def\etal{\emph{et al}\onedot}

\makeatother

\newcommand{\inlinesection}[1]{\vspace{0.05cm} \noindent {\bf #1}}
\newcommand{\titlecaption}[2]{\caption{\textbf{#1.}\xspace#2}}

\newcommand{\titlecaptionof}[3]{\captionof{#1}{\textbf{#2.}\xspace#3}}

\usepackage{pifont}
\newcommand{\cmark}{\ding{51}}%
\newcommand{\xmark}{\ding{55}}%

\usepackage{blindtext}

\definecolor{mpurple}{RGB}{106,27,154}
\definecolor{mpurplelight}{RGB}{206,147,216}

\definecolor{mblue}{RGB}{40,53,147}
\definecolor{mbluelight}{RGB}{159,168,218}

\definecolor{mteal}{RGB}{0,105,92}
\definecolor{mteallight}{RGB}{128,203,196}

\definecolor{morangelight}{RGB}{255,171,145}

\definecolor{mgrayblue}{RGB}{55,71,79}
\definecolor{mgraybluelight}{RGB}{176,190,197}

\definecolor{mamber}{RGB}{255,143,0}
\definecolor{mamberlight}{RGB}{255,224,130}

\definecolor{mdeeporange}{RGB}{216,67,21}
\definecolor{morange}{RGB}{245,124,0}
\definecolor{myellow}{RGB}{253,216,53}

\definecolor{mgreen}{RGB}{85,139,47}
\definecolor{mgreenlight}{RGB}{174,213,129}

\definecolor{mred}{RGB}{198,40,40}
\definecolor{mredlight}{RGB}{239,154,154}

\definecolor{corange1}{RGB}{254,237,222}
\definecolor{corange2}{RGB}{253,190,133}
\definecolor{corange3}{RGB}{253,141,60}
\definecolor{corange4}{RGB}{230,85,13}
\definecolor{corange5}{RGB}{166,54,3}

\definecolor{cblue1}{RGB}{239,243,255}
\definecolor{cblue2}{RGB}{189,215,231}
\definecolor{cblue3}{RGB}{107,174,214}
\definecolor{cblue4}{RGB}{49,130,189}
\definecolor{cblue5}{RGB}{8,81,156}

\definecolor{cgray1}{RGB}{247,247,247}
\definecolor{cgray2}{RGB}{204,204,204}
\definecolor{cgray3}{RGB}{150,150,150}
\definecolor{cgray4}{RGB}{99,99,99}
\definecolor{cgray5}{RGB}{37,37,37}

\definecolor{cred1}{RGB}{254,229,217}
\definecolor{cred2}{RGB}{252,174,145}
\definecolor{cred3}{RGB}{251,106,74}
\definecolor{cred4}{RGB}{222,45,38}
\definecolor{cred5}{RGB}{165,15,21}

\definecolor{cgreen1}{RGB}{237,248,233}
\definecolor{cgreen2}{RGB}{186,228,179}
\definecolor{cgreen3}{RGB}{116,196,118}
\definecolor{cgreen4}{RGB}{49,163,84}
\definecolor{cgreen5}{RGB}{0,109,44}

\definecolor{cdiv11}{RGB}{166,97,26}
\definecolor{cdiv12}{RGB}{223,194,125}
\definecolor{cdiv13}{RGB}{245,245,245}
\definecolor{cdiv14}{RGB}{128,205,193}
\definecolor{cdiv15}{RGB}{1,133,113}

\definecolor{cdiv21}{RGB}{208,28,139}
\definecolor{cdiv22}{RGB}{241,182,218}
\definecolor{cdiv23}{RGB}{247,247,247}
\definecolor{cdiv24}{RGB}{184,225,134}
\definecolor{cdiv25}{RGB}{77,172,38}

\definecolor{cdiv31}{RGB}{230,97,1}
\definecolor{cdiv32}{RGB}{253,184,99}
\definecolor{cdiv33}{RGB}{247,247,247}
\definecolor{cdiv34}{RGB}{178,171,210}
\definecolor{cdiv35}{RGB}{94,60,153}

\definecolor{cdiv41}{RGB}{216,179,101}
\definecolor{cdiv42}{RGB}{245,245,245}
\definecolor{cdiv43}{RGB}{90,180,172}

%% file: figs/teaser.tex
\begin{center}
    \centering
    \captionsetup{type=figure}
    \includegraphics[trim=1 10 0 0,clip,width=\textwidth]{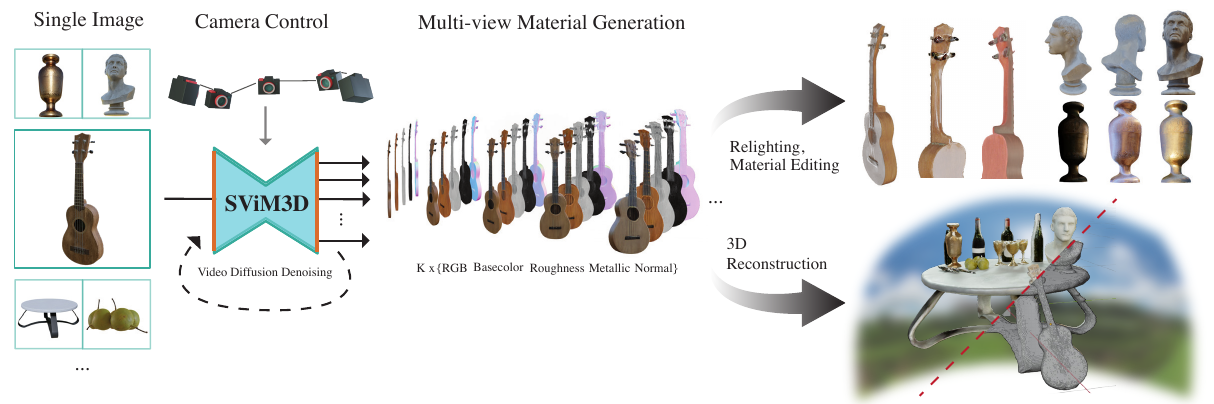}
    \titlecaption{\shortTitle}{\shortTitle~predicts multi-view-consistent spatially-variant material parameters and normals in addition to RGB, conditioned on a single image and a camera path. In addition to relighting and material editing a subsequent optimization stage enables high-quality 3D asset generation for physically based rendering (PBR). Visit the project page at \url{https://svim3d.aengelhardt.com}.
    }
\label{fig:teaser}
\end{center}%

%% file: 00_abstract.tex
\begin{abstract}
We present \longerTitle~(\shortTitle), a framework to predict multi-view consistent physically based rendering (PBR) materials, given a single image. 
Recently, video diffusion models have been successfully used to reconstruct 3D objects from a single image efficiently. However, reflectance is still represented by simple material models or needs to be estimated in additional steps to enable relighting and controlled appearance edits.
We extend a latent video diffusion model to output spatially varying PBR parameters and surface normals jointly with each generated view based on explicit camera control.
This unique setup allows for relighting and generating a 3D asset using our model as neural prior. We introduce various mechanisms to this pipeline that improve quality in this ill-posed setting. 
We show state-of-the-art relighting and novel view synthesis performance on multiple object-centric datasets. 
Our method generalizes to diverse inputs, enabling the generation of relightable 3D assets useful in AR/VR, movies, games and other visual media.
\end{abstract}

%% file: 01_intro.tex
\vspace{-3mm}
\section{Introduction}
\label{sec:intro}
\vspace{-2mm}

\renewcommand{\thefootnote}{\fnsymbol{footnote}}

3D asset generation and relighting are important tasks for various use cases in movies, gaming, e-commerce, and AR/VR. 
In nearly all cases, objects are placed in new environments and lighting conditions. This means illumination information needs to be disentangled from an object's shape and material robustly for it to integrate seamlessly into a new scene. 
Think of the subtle, yet essential differences between a glossy metallic and a matte finish. For generative 3D models without precise material prediction, relighting becomes nearly impossible, resulting in assets that feel out of place. Estimating these properties from a single image under natural illumination, also known as \emph{inverse rendering}~\cite{kaya2021uncalibrated,Nam2018,sang2020single,Zhang2020InverseRendering,Boss2020}, is a highly ill-posed and still unsolved problem.

\vspace{1mm}
\noindent \textbf{Multi-view Material Generation.}
In this work, we present \longerTitle~(\shortTitle), a probabilistic generative diffusion model that tackles object-centric inverse rendering from a single image. Conditioned on a camera pose sequence it generates both high-quality appearances as well as the corresponding multi-view consistent material properties. 
Unlike prior approaches that decouple material estimation from 3D reconstruction, \shortTitle~is the first camera-controllable multi-view model that can produce fully spatially-varying PBR parameters, RGB color and surface normals simultaneously.
The additional output can be leveraged in various applications, hence we consider \shortTitle~ a foundational model that provides a unified neural prior for both 3D reconstruction and material understanding. \shortTitle's output can be used to relight the views directly, perform material edits or to generate full 3D assets by lifting the multi-view material parameters to 3D.
\input{figs/intro_novel_contributions}
As 3D training data paired with material parameters is scarce, we leverage the accumulated world knowledge of a latent video diffusion model~\cite{blattmannStableVideoDiffusion2023}. Specifically, we adapt SV3D~\cite{voletiSV3DNovelMultiview2024}, a video diffusion model~\cite{blattmannStableVideoDiffusion2023} fine-tuned for camera-control by incorporating several crucial modifications:

\begin{itemize}
    \item \textbf{Multi-illumination multi-view training dataset}: We render a high-quality photorealistic synthetic dataset, 
    capturing the complexity of real-world lighting and material variations.
    \item \textbf{Material latent representation}: We treat the material parameters and surface normals as images reusing the image-based autoencoder to encode all inputs into unified latents.
    \item \textbf{Adapted UNet architecture}: We make crucial changes in the core architecture and training scheme to smoothly adapt from image to image+material+normal generation.
\end{itemize}
We use the multi-view PBR video output of SViM3D as pseudo-ground truth for 3D reconstruction. To achieve high-quality 3D reconstructions, we introduce several innovations in our 3D optimization:
\begin{itemize}
    \item \textbf{View-dependent masking}: 
    Loss contributions of the generated views are weighted based on perspective distortion to ensure that material details remain coherent.%
    \item \textbf{Homography correction}: A learnable homography correction mitigates residual multi-view inconsistencies, enhancing reconstruction fidelity.
    \item \textbf{Fast differentiable environment-based lighting}: 
    Our novel differentiable rendering module leverages pre-computed multi-level illumination pyramids to achieve both faster and more accurate lighting optimization.
\end{itemize}

\fig{teaser} highlights examples of relighting and 3D assets in novel environments.
We extensively evaluate \shortTitle~on novel view synthesis (NVS), material prediction, relighting and 3D generation. Our method not only achieves state-of-the-art multi-view consistency but also significantly improves material reproduction in real-world settings 
as the approach inherently understands and exploits multi-view appearance consistency.

\iftoggle{cvprfinal}{
\noindent Please find more examples of our results and more at \url{https://svim3d.aengelhardt.com}.
}

%% file: figs/intro_novel_contributions.tex
\begin{figure}[tb]
    \centering
\resizebox{.90\linewidth}{!}{
\begin{tikzpicture}[
    node distance=0.2cm,
    font={\fontsize{8pt}{10}\selectfont},
    spy using outlines={circle, magnification=1.5, connect spies}
]
\node[] (input_masking) {\includegraphics[trim={2.5cm 2.5cm 2.5cm 2.5cm},clip,width=2cm]{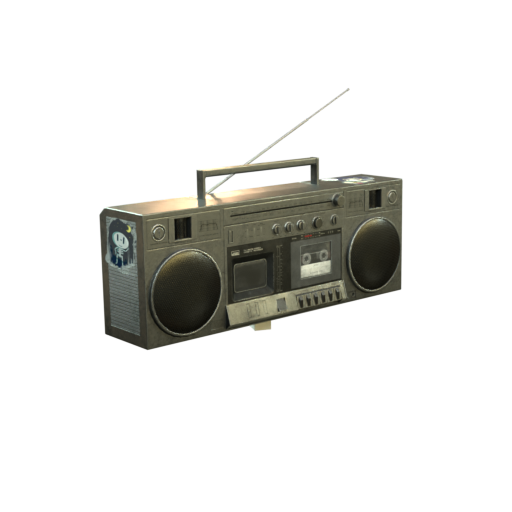}};
\node[right=of input_masking] (no_masking) {\includegraphics[trim={2.5cm 2.5cm 2.5cm 2.5cm},clip,width=2cm]{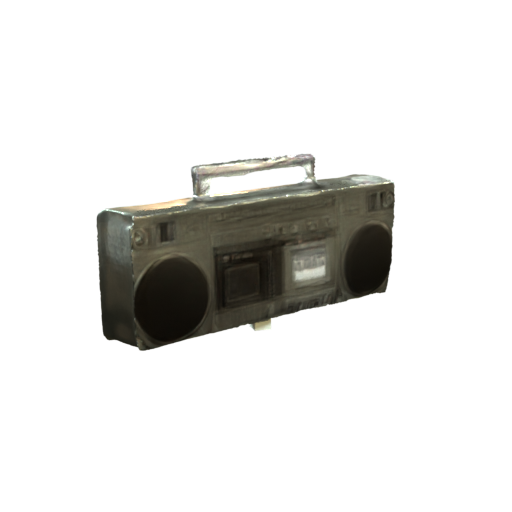}};
\spy[size=0.85cm] on ($(no_masking)+(0.65cm,0.25cm)$) in node at ($(no_masking.south)+(0.5cm,0.3cm)$);
\node[right=of no_masking] (ours_masking) {\includegraphics[trim={2.5cm 2.5cm 2.5cm 2.5cm},clip,width=2cm]{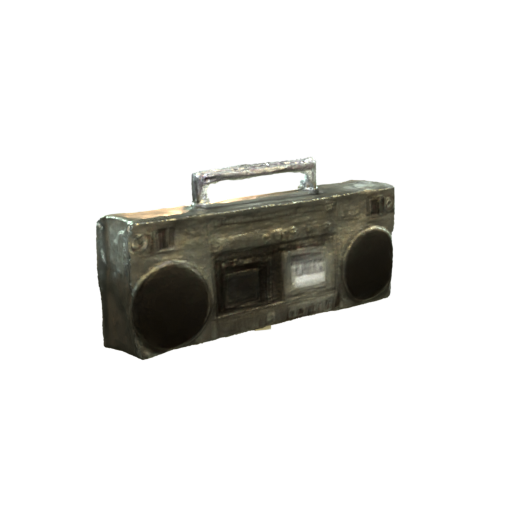}};
\spy[size=0.85cm] on ($(ours_masking)+(0.65cm,0.25cm)$) in node at ($(ours_masking.south)+(0.5cm,0.3cm)$);

\node[above=-0.2cm of input_masking] (input_label) {GT};
\node[above=-0.2cm of no_masking] (no_label) {w/o};
\node[above=-0.2cm of ours_masking] (ours_label) {\shortTitle~(Ours)};
\node[left=-0.2cm of input_masking.west, rotate=90, anchor=south, text width=1.0cm, align=center] (masking_label) {\baselineskip=8pt View Masking \par};

\node[below=-0.2cm of input_masking] (input_correction) {\includegraphics[trim={1cm 0cm 1cm 1cm},clip,width=2cm]{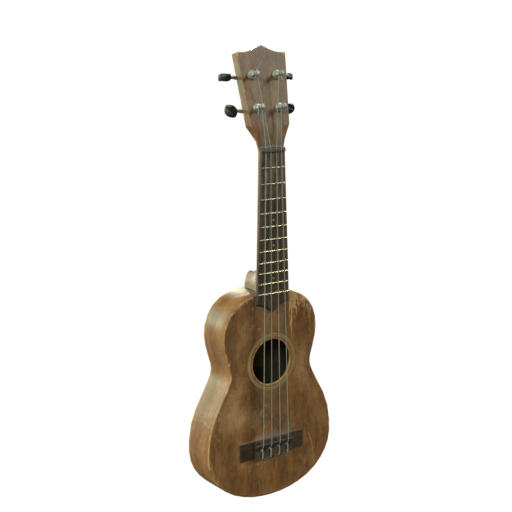}};
\node[right=of input_correction] (no_correction) {\includegraphics[trim={1cm 0cm 1cm 1cm},clip,width=2cm]{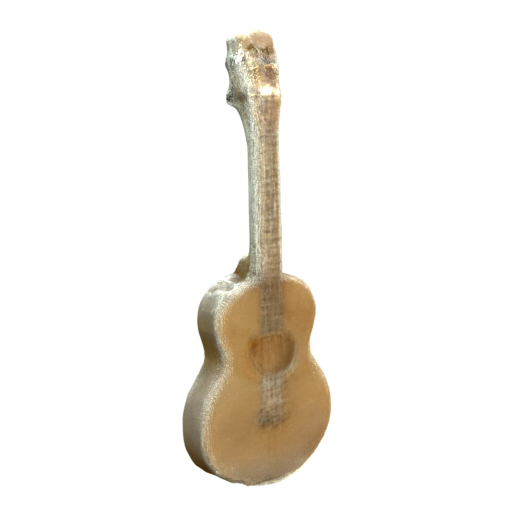}};
\spy[size=0.85cm] on ($(no_correction)+(-0.05cm,0.00cm)$) in node at ($(input_correction.east)+(0.1cm,0cm)$);
\node[right=of no_correction] (ours_correction) {\includegraphics[trim={1cm 0cm 1cm 1cm},clip,width=2cm]{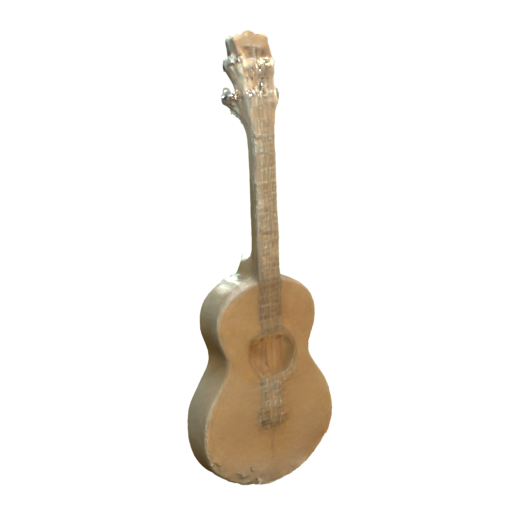}};
\spy[size=0.85cm] on ($(ours_correction)+(-0.05cm,0.00cm)$) in node at ($(no_correction.east)+(0.1cm,0cm)$);
\node[left=0cm of input_correction.west, rotate=90, anchor=south, text width=1.2cm, align=center] (correction_label) {\baselineskip=8pt Homography correction \par};

\node[below=-0.2cm of input_correction] (input_lighting) {\includegraphics[width=2cm]{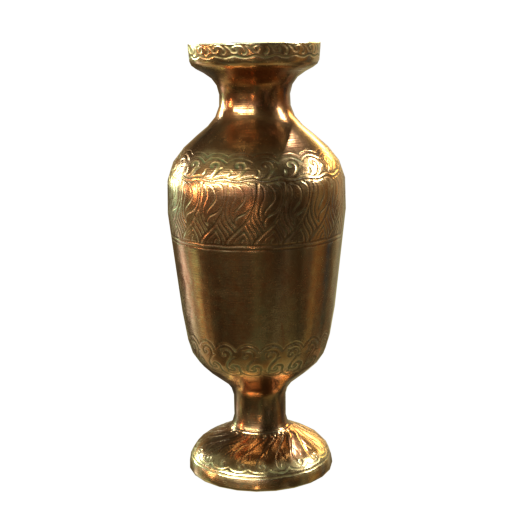}};
\node[right=of input_lighting] (no_lighting) {\includegraphics[width=2cm]{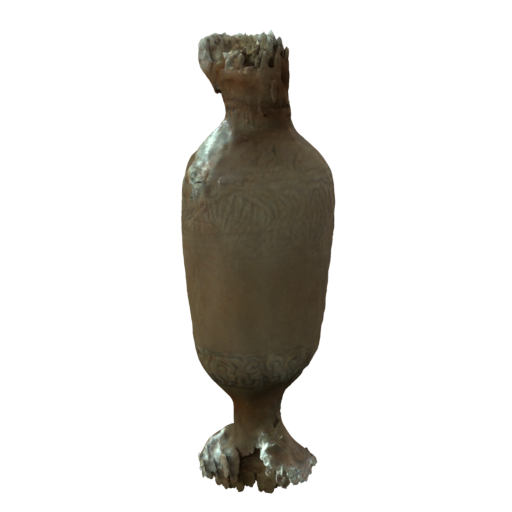}};
\spy[size=0.85cm] on ($(no_lighting)+(-0.15cm,0.25cm)$) in node at ($(input_lighting.east)+(0.1cm,0cm)$);
\node[right=of no_lighting] (ours_lighting) {\includegraphics[width=2cm]{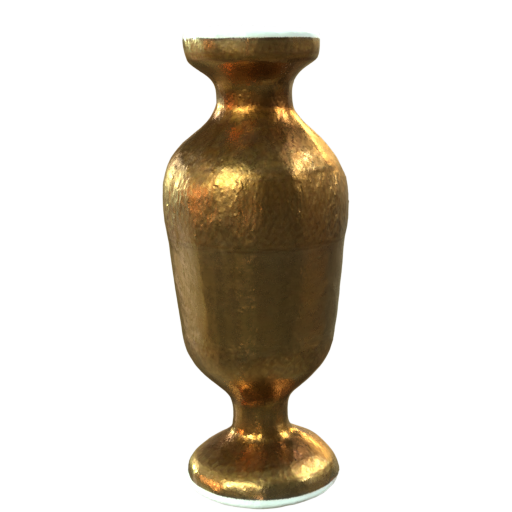}};
\spy[size=0.85cm] on ($(ours_lighting)+(-0.15cm,0.25cm)$) in node at ($(no_lighting.east)+(0.1cm,0cm)$);
\node[left=0cm of input_lighting.west, rotate=90, anchor=south, text width=1.2cm, align=center] (correction_label) {\baselineskip=8pt Environment Lighting \par};

\end{tikzpicture}%

}
\vspace{-.75em}
\titlecaption{\shortTitle~Improvements on Common Issues}{%
Our method introduces several new contributions which improve the reconstruction quality of our method drastically.
}
\label{fig:issues}
\vspace{-.75em}
\end{figure}

%% file: 02_related.tex
\vspace{-2mm}
\section{Related works}
\label{sec:related}
\vspace{-2mm}

Inverse rendering is a challenging and ambiguous problem, traditionally performed in controlled laboratory settings~\cite{Asselin2020, Boss2018, lawrence2004, Lensch2001, Lensch2003, xuHybridMeshneuralRepresentation2022}.
Building on insights from constrained estimation, 
various methods propose casual acquisition setups for
planar surfaces, using single shot~\cite{Aittala2018, Deschaintre2018, henzler2021, Li2018, sang2020single, Boss2019}, few-shot~\cite{Aittala2018,XuMinimalBRDFSampling2026} or multi-shot~\cite{Albert2018, Boss2018, Deschaintre2019, Deschaintre2020,Gao2019} captures.
Casual capture has also been extended to joint lighting model and shape reconstruction~\cite{bi2020a,bi2020b,bi2020c,Boss2020,kaya2021uncalibrated,Nam2018,sang2020single,Zhang2020InverseRendering}, even on scenes~\cite{li2020inverse,Sengupta2019}.
Recovering lighting under unknown passive illumination is significantly more challenging as it requires disentangling shape and materials from the illumination.

\inlinesection{Implicit representations.} 
Methods based on neural fields achieved decomposition of scenes under varying illumination~\cite{Boss2021, Boss2021neuralPIL} or fixed illumination~\cite{zhang2021, liangENVIDRImplicitDifferentiable2023,zhang2023neilf++, nerfactor_zhang21,zhang2022invrender}, even with uncertain or unknown camera parameters~\cite{bossSAMURAIShapeMaterial2022,engelhardt2023-shinobi}. Also, 3D Gaussians have been explored as scene representation in this context~\cite{gao2023relightable,saito2024rgca}. 
However, all these methods rely on multi-view input and need to be optimized per object.

\inlinesection{3D reconstruction with material prediction.}
BRDF parameter autoencoders~\cite{Boss2021neuralPIL, xuMATLABERMaterialAwareTextto3D2023} or lighting constraints~\cite{Boss2021neuralPIL, gardner2022rotationequivariant, gardner2023renipp} have shown to help with inverse rendering. Recently, diffusion models have gained traction for their probabilistic handling of ambiguity in casual capture scenarios. Du \etal~\cite{duGenerativeModelsWhat2023} explore intrinsic imaging with diffusion models, leveraging LoRA~\cite{hu2022lora} and small datasets, showing Stable Diffusion’s~\cite{rombachHighResolutionImageSynthesis2022} potential, despite quality limits. Material Palette~\cite{lopesMaterialPaletteExtraction2023a} and ControlMat~\cite{vecchioControlMatControlledGenerative2023} generate tileable textures, while Xu \etal~\cite{xuMATLABERMaterialAwareTextto3D2023} incorporate SDS loss~\cite{poole2023dreamfusion} and Deep Marching Tetrahedra (DMTet)~\cite{munkberg2022nvdiffrec,shen2021dmtet} for reconstruction. Intrinsic Image Diffusion (IID)~\cite{kocsisIntrinsicImageDiffusion2023} is one of the first works to explore diffusion models for PBR material estimation, it fine-tunes Stable Diffusion on an interior dataset~\cite{zhu2022learninginterior} for PBR parameter prediction and relighting. RGB$\leftrightarrow$X~\cite{zengRGBX2024} estimates PBR data as part of their material- and lighting-aware neural rendering pipeline, their model can predict either albedo, roughness, metallic or diffuse irradiance maps conditioned on a single image and a text prompt.
MaterialFusion~\cite{litmanMaterialFusionEnhancingInverse2024} proposes a 2D material denoising diffusion prior called StableMaterial based on StableDiffusion 2.1~\cite{rombachHighResolutionImageSynthesis2022} but trained on object centric data, and employs an SDS-based optimization to achieve 3D asset generation. Gaussian-ID~\cite{duGSIDIlluminationDecomposition2024} proposes 3D reconstruction with diffusion-based material priors, using multi-view data and 3D Gaussian Splatting~\cite{kerbl3Dgaussians} with parametric lighting, building on Kocsis \etal~\cite{kocsisIntrinsicImageDiffusion2023}.
In contrast to all these methods, SViM3D jointly estimates all material parameters in a multi-view consistent manner,
making it a robust foundation model for 3D reconstruction.

\inlinesection{3D generation with materials.} 
In contrast to texture generation given 3D geometry~\cite{collaborative_control_vainer_2024,deng2024flashtex,youwang2024paintit,zeng2024paint3d} we focus on joint 3D and material generation. Recent techniques in 3D generation often guide the optimization of DMTet~\cite{munkberg2022nvdiffrec} with a reflectance field. A special case of these 3D generations is to create an asset from a single image. These approaches~\cite{hongLRMLargeReconstruction2023, voletiSV3DNovelMultiview2024, clay, siddiquiMeta3DAssetGen, boss_sf3d2024, huang2025spar3d} benefit from large-scale pretraining on image data, often followed by a supervised fine-tuning on synthetic data.
First steps in diffusion based pair-wise view generation with camera control have been achieved by zero-123~\cite{liu2023zero} and its follow-ups.
Recently, video data has also been utilized in the context of video diffusion models~\cite{chenV3DVideoDiffusion2024a,voletiSV3DNovelMultiview2024}. Guidance from a pretrained image/video diffusion model is leveraged using either (1) Score Distillation Sampling (SDS)~\cite{poole2023dreamfusion} loss, or (2) photometric reconstruction loss. 
Fantasia3D~\cite{chenFantasia3DDisentanglingGeometry2023} uses Stable Diffusion~\cite{rombachHighResolutionImageSynthesis2022}, UniDream~\cite{liu2023unidream} predicts normals and albedo with a multi-view diffusion model. RichDreamer~\cite{qiuRichDreamerGeneralizableNormalDepth2023} employs two models for albedo and normal-depth generation. However, SDS optimization has several drawbacks, including multi-view inconsistency, long runtimes, and issues with oversaturated colors and blurry details.
AssetGen~\cite{siddiquiMeta3DAssetGen} replaces SDS-based optimization with a photometric loss, using a multi-view diffusion model for albedo and radiance prediction. A transformer converts views into a triplane representation~\cite{hongLRMLargeReconstruction2023}, enabling 3D reconstruction with UV texture refinement.
CLAY~\cite{clay} generates materials conditioned on geometry previously generated by a different module using an adapted MVDream~\cite{shiMVDreamMultiviewDiffusion2024} model without explicit camera control.
However, PBR parameters that enable relighting still require multi-view data input with potentially changing illumination~\cite{Boss2021neuralPIL, gao2023relightable}, additional workflow steps or text input~\cite{siddiquiMeta3DAssetGen,clay,mercierHexaGen3DStableDiffusionJust2024,chenFantasia3DDisentanglingGeometry2023}, or are simply not available in current methods~\cite{chenV3DVideoDiffusion2024a,voletiSV3DNovelMultiview2024,huangZeroShapeRegressionbasedZeroshot2024}.
Recent work has shown that direct image-to-image relighting without explicit 3D reconstruction is feasible using diffusion-based models conditioned on illumination~\cite{jin2024neural_gaffer, zeng2024dilightnet, zhang2025scaling_iclight}.
In contrast, \shortTitle predicts RGB, PBR parameters, and normals simultaneously, enabling explicit 3D shape and illumination reconstruction using a single model and achieving high 3D consistency.

%% file: 03_method.tex
\input{figs/overview}
\vspace{-2mm}
\section{Preliminaries}
\vspace{-2mm}

\inlinesection{Video diffusion based 3D generation.}
Recently, video diffusion models have been exploited for novel view synthesis and 3D reconstruction tasks~\cite{chenV3DVideoDiffusion2024a,voletiSV3DNovelMultiview2024}, due to improved view consistency and generalization from being trained on huge image and video datasets~\cite{schuhmann2022laionb, xiong2024lvd2m}. A 3D video diffusion model is conditioned on a reference image $\mathbf{I} \in \mathbb{R}^{3 \times H \times W}$ of an object, and a camera orbit of a sequence of $K$ poses around it. It is then trained to generate a $K$-frame orbital video $\mathbf{M} \in \mathbb{R}^{K\times3\times H \times W}$ around the object using the diffusion formulation~\cite{song2019,ho2020denoising,voleti2022mcvd,blattmannStableVideoDiffusion2023,voleti2024sv3d}.
During inference, the user provides an image and camera trajectory, and the 3D video diffusion model is used to iteratively generate an orbital video in multiple diffusion steps.
These views can then be used in a multi-view 3D reconstruction pipeline~\cite{munkberg2022nvdiffrec,mildenhall2020,kerbl3Dgaussians,voleti2024sv3d}.

\inlinesection{Physically-based rendering (PBR).}
From a rendering perspective, an object is rendered as an image by computing its radiance at each location, which is then denoted by its RGB pixel value $\vect{c} \in \mathbb{R}^3$. Radiance is computed by appropriately factoring in the contributions of the object's PBR material properties $\vect{b} := [\vect{b}_c; b_r; b_m]$ consisting of albedo basecolor $\vect{b}_c \in \mathbb{R}^3$, roughness $b_r \in \mathbb{R}$, and metallic-ness $b_m \in \mathbb{R}$; as well as its surface normal $\vect{n} \in \mathbb{R}^3$.
Specifically, the outgoing radiance $\vect{c} = L(\vect{\omega_o})$ in direction $\vect{\omega_o}$ is defined by a simplified rendering equation~\cite{Kajiya1986} as:
\vspace{-.5em}
\begin{equation}
    L(\vect{\omega_o})=\int_\Omega L_i(\vect{\omega_i}) f(\vect{\omega_i},\vect{\omega_o}) (\vect{\omega_i} \cdot \vect{n}) d\vect{\omega_i},
    \label{eq:rendering_equation}
\vspace{-.5em}
\end{equation}
\ie the integral over the hemisphere $\Omega$ of the %
incoming light $L_i(\vect{\omega_i})$ from direction $\vect{\omega_i}$ multiplied with the Bidirectional Reflectance Distribution Function (BRDF) $f(\vect{\omega_i}, \vect{\omega_o})$ and the cosine shading term $(\vect{\omega_i} \cdot \vect{n})$. We model the specular portion of the BRDF with the analytical Cook-Torrance microfacet model~\cite{Cook1982}, yielding: 
\vspace{-.5em}
\begin{equation}
    f(\vect{\omega_i},\vect{\omega_o}) = \frac{D F G}{4 (\vect{\omega_o} \cdot \vect{n}) (\vect{\omega_i} \cdot \vect{n})} + \frac{\vect{b}_c}{\pi}(1-b_m)
\label{eq:brdf}
\vspace{-.5em}
\end{equation}
where $D$, $F$, $G$ represent the normal distribution function (NDF), Fresnel term, and the geometric attenuation function, respectively. For $D$ we rely on the GGX distribution~\cite{Walter2007}. We adopt the parametrization of the Disney BRDF~\cite{Burley2012}, where $\vect{b}_c$ and $b_m$ characterize $F$, and $b_r$ characterizes $D$ and $G$~\cite{pharrPhysicallyBasedRendering2017,VeachPhD}.
In \shortTitle, we adapt a 3D video diffusion model framework~\cite{voleti2024sv3d}, and jointly generate images $\vect{c}$, PBR materials $\vect{b}$, and normal $\vect{n}$ for each target view.

\vspace{-2mm}
\section{\shortTitle: Multi-view PBR Generation}
\label{sec:method}
\vspace{-2mm}

\inlinesection{Overview.}
The aim of \shortTitle~is to convert a single 2D image and a camera orbit into RGB frames, corresponding material parameters, and normal maps. Fig.~\ref{fig:overview} lays out the key components of our method.

\inlinesection{Problem setup.}
The inputs to the model are:
\begin{itemize}
\item A color image $\mathbf{I} \in \mathbb{R}^{3 \times H \times W}$ of an object, and
\item A camera pose trajectory defined by tuples of elevation and azimuthal angles $\vect{\pi} \in \mathbb{R}^{K\times 2} = \{(e_i, a_i) \}^K_{i=1}$ centered at the object, with $K=21$ views. 
\end{itemize}
The goal is to estimate an augmented orbital video $\mathbf{M} \in \mathbb{R}^{K\times11\times H \times W}$, \ie a $K$-frame video of 11-channel frames:
\begin{itemize}
\item 3 channels for the image RGB color $\vect{c} \in \mathbb{R}^3$,
\item 5 channels for PBR parameters from the Cook-Torrance model~\cite{Cook1982} $\vect{b} \in \mathbb{R}^5$, namely basecolor $\vect{b}_c \in \mathbb{R}^3$, roughness $b_r \in \mathbb{R}$, metallic $b_m \in \mathbb{R}$, and
\item 3 channels for the unit-length surface normal in camera space $\vect{n} \in \mathbb{R}^3$.
\end{itemize}
\shortTitle~is trained to iteratively generate $\mathbf{M}$ through a denoising diffusion process, similar to 3D video diffusion models but with the above augmented channels.

\subsection{SViM3D Architecture and Training}
The architecture of \shortTitle~is built on that of SV3D~\cite{voletiSV3DNovelMultiview2024}, which in turn is built on that of SVD~\cite{blattmannStableVideoDiffusion2023}. However, we introduce important elements to the architecture in order to adapt to generating material parameters and normals.

\inlinesection{Material latent representation.}
While it is prudent to re-use a pretrained 3D video diffusion model to leverage the rich image and video priors it has learned, such models operate in the latent space, using a variational autoencoder (VAE) to first encode the images into latents, perform a denoising step, then finally decode latents into images. Material properties or normals do not have a standard latent representation, though.
Inspired by other diffusion model works~\cite{ke2023repurposing,zhang2023adding} that take additional conditioning, our main insight is that the VAE of an image generative model can encode material properties and surface normals, by treating them as images.

The VAE of SV3D takes a 3-channel image input, and outputs a 4-channel latent at $1/8$th the original image side-length. For training, we use this VAE to encode the RGB image $\vect{c}$; albedo basecolor $\vect{b}_c$; a concatenation of roughness and metallic-ness padded with zeros to make 3 channels $[0; b_r; b_m]$ to align with the Occlusion-Roughness-Metallic (ORM) channel layout often used in real-time graphics~\cite{GLTF_manual}; and the surface normal $\vect{n}$, each into 4-channel latents. Therefore, the network predicts a 16-channel stack. We preprocess all latents, and feed the concatenated tensors to the UNet after adding the time step-specific noise.

\inlinesection{UNet adaptation.}
Each denoising step is performed by a UNet with multiple layers at different scales. Each layer consists of one residual block with Conv3D layers, and two transformer blocks (spatial and temporal) with attention layers as illustrated in Fig~\ref{fig:overview}. While SV3D captures only the latents of the RGB frames of the orbital video, we augment the input and output to include PBR material and surface normal by extending the channel dimension of the input and output layers from 4 to 16. The newly extended weights are initialized by repeatedly copying the existing parameters from the weights for the RGB latent channels.
The rest of the architecture follows that of SV3D~\cite{voleti2024sv3d}.

\inlinesection{Multi-illumination multi-view training dataset.}
We combine multiple data sources and render a synthetic photorealistic dataset using Blender's Cycles~\cite{blender} render engine. We exclude data that includes subsets of the Poly Haven data which we use for testing.
Per object, we randomly select four environment maps. %
For each illumination setting, we sample a random camera trajectory with a fixed distance between camera and object that also ensures that the convex hull of the scene content is inside the camera frustum for all views.

\inlinesection{Training details.} 
We use the popular EDM framework~\cite{karras2022elucidating} for training with the simplified diffusion loss from~\cite{blattmannStableVideoDiffusion2023}. 
While the pre-trained VAE is reused, we train the denoising UNet in two phases. First, we freeze all temporal attention blocks (see Fig.~\ref{fig:overview}) while training on all data for roughly 100k steps. Examples featuring low quality material parameters like missing texture maps or uniform parameters we only use for RGB and normal supervision.
Afterwards, we finetune the whole UNet on the highest quality PBR data for another 60k steps. This staged training helps avoid forgetting of the temporal knowledge in the initial model, and stabilizes the training for task adaptation compared to a full fine-tuning from the start.  
For inference, we follow a triangular Classifier-free Guidance (CFG)~\cite{ho2021classifierfree} scaling similar to that of SV3D~\cite{voletiSV3DNovelMultiview2024}.

\subsection{Relighting}
\label{sec:relighting}
Adjusting an object's appearance to fit into a new environment is critical for seamless integration into AR or believable compositions in media production.
\shortTitle~generates material parameters that can be directly used for (re-)lighting in image space without an explicit surface reconstruction (2.5D). Given an illumination representation like an HDRI and a virtual camera we can use the generated normal direction to define the ray geometry to evaluate the predicted BRDF. 
Using a split sum illumination model (Eq.~\ref{eq:split_sum}) we can compute the lighting at interactive rates (see ~\cref{fig:res_mv_materials}).

\inlinesection{Fast environment-based lighting.} %
We propose a fast environment map-based rendering, which can encode significantly more lighting details than the low-frequency illumination models used in SV3D~\cite{voletiSV3DNovelMultiview2024}, for example. Our image-based lighting~\cite{Karis2013} which leverages pre-filtered importance sampling~\cite{Krivanek2008prefilteredIS} for fast integration of the incoming light supports both 2.5D relighting in image space as well as 3D reconstruction and rendering. The third row of \fig{issues} shows that our scheme delivers better image-based lighting.

Specifically, as Monte Carlo integration is costly and can lead to high noise levels, we adopt the split sum approximation~\cite{Karis2013} from real-time rendering. This technique has proven effective in prior work~\cite{Boss2021neuralPIL, munkberg2022nvdiffrec}, and approximates the integral of \eqn{rendering_equation} as:

{\footnotesize
\begin{equation}
    L(\vect{\omega_o})=\int_\Omega f(\vect{\omega_i},\vect{\omega_o}) (\vect{\omega_i} \cdot \vect{n}) d\vect{\omega_i} \int_\Omega L_i(\vect{\omega_i}) D(\vect{\omega_i},\vect{\omega_o}) (\vect{\omega_i} \cdot \vect{n}) d\vect{\omega_i}
\label{eq:split_sum}
\end{equation}
}
The first integral depends only on BRDF roughness and the cosine term, and is pre-computed into a 2D lookup texture. The second term, involving incoming radiance and the NDF $D$, is pre-integrated into a filtered environment map at multiple fixed roughness levels. Since rougher materials need lower resolution, we store the result in an image pyramid, or \emph{environment pyramid}. Rendering then becomes a multiplication of two lookups based on $(r,(\vect{\omega_i} \cdot \vect{n}))$ and a pyramid level selected by $r$ and direction $\vect{\omega_i}$.  
To account for multiple scattering, we use attenuated cosine-weighted radiance from a lower mip level, following~\cite{Carmelo2019_multiple_scattering}.

As precomputing the filtered environment map is expensive, we introduce several optimizations. Unlike~\cite{munkberg2022nvdiffrec}, we use Monte Carlo integration. 
For increasing roughness values from $[0-1]$ we use 0, 4, 16, 24, and 24 samples over 5 \textit{mip} levels during optimization, and 8 levels with up to 256 samples for relighting. The first level (mirror direction) is excluded. To reduce noise, we apply filtered importance sampling~\cite{Krivanek2008prefilteredIS}, where environment resolution is adapted to sample likelihood. For diffuse lighting, we filter a low-res environment image (no NDF) using 16 samples. We reduce the perceptible noise by drawing random samples from the Halton sequence. While our method supports arbitrary environment map formats, we find octahedral maps~\cite{Engelhardt2008OctahedronEM} to work well as they yield fewer pole artifacts than spherical ones. Rendering is performed in linear HDR color space and tonemapped using AgX~\cite{SobotkaAgX}.

\subsection{3D Reconstruction using~\shortTitle~Outputs}
\label{sec:3d_reconstruction}
We use the outputs of \shortTitle~as pseudo-ground-truth (pGT) for 3D reconstruction. Our pipeline is agnostic to the 3D representation, we use a NeRF-based implicit function. For illumination we use our environment lighting representation introduced above. It is implemented in pure PyTorch and, therefore, fully differentiable while still fast enough to execute the pre-filtering in each training step.
As visualized in Fig.~\ref{fig:overview}, our reconstruction pipeline comprises four phases: 
\begin{enumerate}\addtocounter{enumi}{-1}
    \item An illumination representation is pre-optimized using the orbital video $\mathbf{M}$ to initialize Phase 1.
    \item A modified Instant-NGP~\cite{Mueller2022} is optimized using a photometric rendering loss relying on the jointly optimized illumination and supervision from the reference views.
    \item We optimize a DMTet~\cite{munkberg2022nvdiffrec} representation initialized from the results of Phase 1 via marching cubes.
    \item A mesh is finally extracted, UV unwrapped using xatlas~\cite{youngJpcyXatlas2024} and all textures baked.
\end{enumerate}
The generated asset can be used in any computer graphics pipeline, e.g. integrated into new scenes and lighting.

\inlinesection{Optimization.}
For optimal results, we run Phase 0 for 200 steps, followed by 800 steps of Phase 1, and 1500 steps of Phase 2.
3D reconstruction quality is dependent on the quality of the 3D representation and the illumination. However, lighting effects are highly view-dependent, and cannot be modeled with low-frequency illumination models, leading to degraded reconstruction performance as shown in the last row of \fig{issues}.
We weigh the normal supervision losses high as our pGT are of high quality. For quick previews, Phase 0 and potentially also Phase 2 and Phase 3 can be omitted, giving a 3D representation suitable for novel view synthesis and basic relighting.
In addition to the reconstruction loss, we employ two LPIPS instances on randomly sampled triplets of the rendered output $\hat{\mathbf{I}}$ and the pGT $\mathbf{M}$ inspired by~\cite{chambonPassingMultiChannelMaterial2021a}.
Further details about the losses and optimization process are available in the appendix.

We observe that even slight multi-view inconsistency in the \shortTitle~pGT outputs may lead to blurry results in the 3D asset, as shown in \fig{issues}. Therefore, we introduce \textit{view-dependent masking} and \textit{homography correction} to compensate for minute inconsistencies in the input pGT, and improve the detail drastically. To further add consistency between the rendered result and our PBR prediction, we perform \textit{fast differentiable environment-based lighting} to enable more high-frequency lighting details compared to parametric illumination models~\cite{Boss2021, voletiSV3DNovelMultiview2024}, which leads to textures with even lesser light baked in.

\inlinesection{View-dependent masking.}
We observe that artifacts are prominent in parts of the generated views where the perspective distortion is heaviest.
Therefore, regions in an image with the least distortion should contribute the most. The first row of \fig{issues} shows that our scheme produces more detailed texture information.
A good proxy to identify the trusted areas that are parallel to the image plane is the dot product $\hat{\vect{n}}\cdot\vect{v}_i$ of the bilaterally filtered surface normal $\hat{\vect{n}}$ and view direction $\vect{v}_i$ from surface points $p \in \mathbb{R}^3$ to camera position $\pi_i$. Higher values, i.e. higher correlation between $\vect{v}_i$ and $\hat{\vect{n}}$ indicate better alignment and consequently higher trust. $\mathbf{A}_i$ for view $i$ is then used to mask all geometry related losses.

\inlinesection{Homography correction.}
To address remaining inconsistencies between the pseudo-ground-truth (pGT) views,
we introduce a learnable per-view homography correction. The second row of \fig{issues} shows that our scheme corrects view inconsistencies.
Specifically, for each pGT view, we jointly optimize a homography matrix $H_i$, initialized as the identity transform, for the latter part of Phase 1.
During optimization, $H_i$ is applied to the rendered view $\hat{\mathbf{I}}_i$ during loss computation as $\hat{\mathbf{I}_i}' = H_i\hat{\mathbf{I}}_i$, enabling the model to match the view to the potentially imperfect pGT.

%% file: figs/overview.tex
\begin{figure*}[t]
\begin{center}
\includegraphics[trim=0 185 0 0,clip,width=\textwidth]{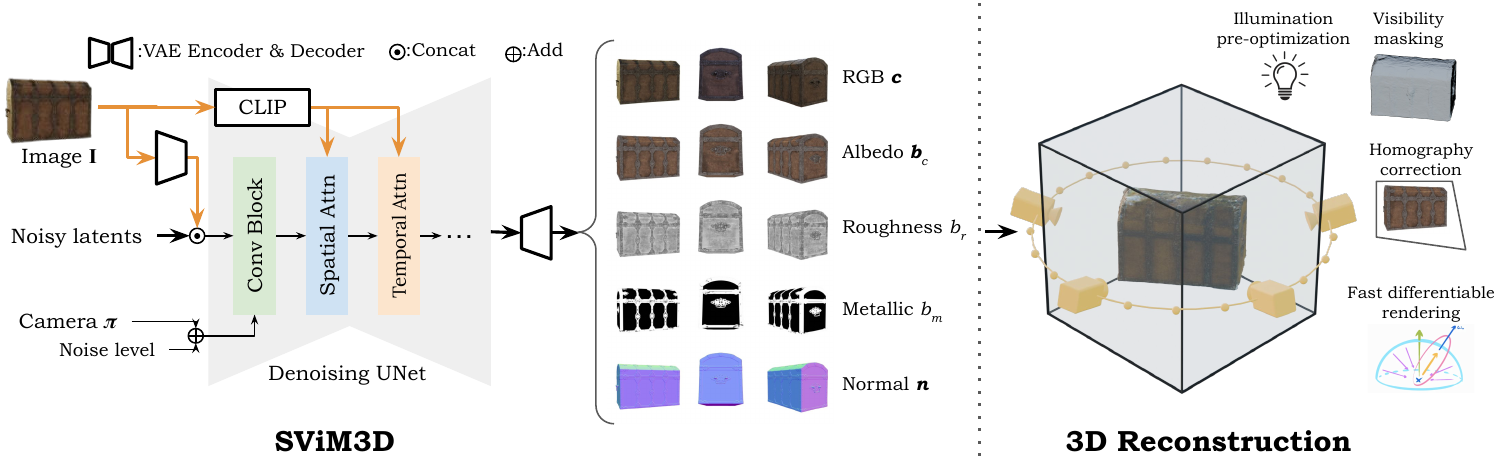}
\end{center}
\vspace{-.75em}
\titlecaption{The \shortTitle~pipeline}{We train a video diffusion model on multi-view and multi-illumination data to generate multi-view images with material parameters. During inference, given a single image, \shortTitle~ can generate 21 views with consistent RGB radiance, albedo, roughness, metallic, and camera space normals. We then use the synthesized novel views for 3D reconstruction that yields textured meshes with PBR materials. Starting from illumination pre-optimization, we further propose several techniques to aid the 3D reconstruction pipeline in this sparse view setting, such as visibility masking, homography correction, fast differentiable rendering. %
}
\label{fig:overview}
\vspace{-.75em}
\end{figure*}

%% file: 04_results.tex
\input{tables/res_single_material}

\section{Experiments and Results}
\label{sec:results}
\vspace{-2mm}

\inlinesection{Datasets.} 
To evaluate our new task set, we introduce the Poly Haven object dataset~\cite{polyhavenPolyHaven} consisting of 315 high-quality assets with PBR materials. 
We render synthetic scenes similar to our training dataset, with 21 frames per orbit and multiple illumination settings per object.
We also evaluate the relighting and 3D reconstruction performance on the Stanford Orb~\cite{kuang2023stanfordorb} benchmark, which contains real-world multi-view scenes with scanned meshes and ground truth environment maps. The supplements contain results from other datasets.

\input{figs/res_mv_comparison}
\input{figs/res_multiview_materials}
\input{figs/res_single_material}

\inlinesection{Metrics.}
The generated RGB radiance and, as an indicator of decomposition quality, the albedo basecolor are evaluated using PSNR, SSIM, and the distribution matching metrics LPIPS~\cite{zhang2018perceptual}, CLIP-Score (CLIP-S)~\cite{hesselCLIPS2021}, the CLIP Maximum-Mean Discrepancy (CMMD)~\cite{Jayasumana_2024_CMMD}, and FID~\cite{Bynagari2019GAN_FID}.
CMMD compares the distribution of the CLIP~\cite{pmlr-v139-radford21a} embeddings of generated and reference images, and has been shown to be a better indicator of low-level image quality than FID~\cite{Jayasumana_2024_CMMD,vainerCollaborativeControlGeometryconditioned2024a}.
As these image metrics have limited meaning for the material maps, we evaluate them via PSNR and root mean squared error (RMSE).
We average over three samples and match the scale and shift for albedo and roughness predictions to the ground truth (GT) to compensate for the inherent ambiguity. 
See the appendix for a visual comparison across multiple samples.
3D optimization is evaluated in the appendix, too.

\inlinesection{Baselines.} 
We compare \shortTitle~with IID~\cite{kocsisIntrinsicImageDiffusion2023}, StableMaterial (SM) of MaterialFusion~\cite{litmanMaterialFusionEnhancingInverse2024}, and RGB$\leftrightarrow$X~\cite{zengRGBX2024}.
Although not capable of PBR material estimation, we also compare against SV3D~\cite{voletiSV3DNovelMultiview2024} which generates multi-view RGB images from a single view, and also allows for 3D reconstruction similarly to our method. 
Since none of the available methods exactly matches our task, \ie joint multi-view and material prediction from a single image with camera control, we compare the models on multiple tasks to evaluate efficacy.
\input{tables/res_mv_rgb_albedo}
\input{tables/ablation_stanford_orb}

\inlinesection{Single image material prediction.}
Given the lack of closely related baselines, we also compare existing techniques on single-image material estimation. Tab.~\ref{tab:single_material} shows the performance for this subtask (on the reference frame) of our task as the condition view is part of the camera trajectory. Results clearly show that we comfortably outperform the single-image material prediction baselines.

\inlinesection{Multi-view novel view image and albedo synthesis (NVS).} NVS is evaluated in the top part of~\cref{tab:mv_rgb_albedo} using the RGB radiance output. The lower part of the table compares multi-view albedo generation, which we obtain for the other methods by first running multi-view NVS for RGB images using SV3D~\cite{voleti2024sv3d}, and then the respective material generation conditioned on each multi-view image. 
We also replicate the GT RGB input frames using the GT illumination and our generated PBR materials with 2.5D relighting (also see Fig.~\ref{fig:res_mv_materials}). Fig.~\ref{fig:app_25d_relighting_comparisons} shows examples for multiple view and light directions, comparing our results to the ground truth and multiple diffusion based relighting baselines. Please find more information on the baseline models and visual examples in the supplements. 
We use the default inference configuration for all other methods, and we use 50 steps of the deterministic DDIM sampler~\cite{DDIM2020} with the guidance scheme described in Sec.~\ref{sec:method} for ours. 
From~\cref{tab:mv_rgb_albedo}, we see that
\shortTitle~achieves state-of-the-art performance on almost all metrics. 
Interestingly, the multi-view RGB prediction improved compared to the baseline, indicating that, although a potentially more challenging task, PBR generation also helps RGB generation.

\inlinesection{3D Relighting and Novel View Synthesis}
We compare rendered 3D reconstruction results using the provided test views with original and novel illuminations from the Stanford Orb benchmark~\cite{kuang2023stanfordorb} in Tab.~\ref{tab:stanford_orb}. The metrics show that the reconstructed material parameters are well suited for physically based rendering in diverse illumination settings and allow better reproduction of the original illumination compared to SF3D~\cite{boss_sf3d2024}. Please find more info on SF3D in the appendix.

\inlinesection{Visual results.}
Fig.~\ref{fig:res_mv_comparison} visualizes the superior 3D consistency of our multi-view generation compared to all other variants. This underlines the benefit of a combined neural prior for video and 3D based applications. The results in Tab.~\ref{tab:single_material} and Tab.~\ref{tab:mv_rgb_albedo} support these observations. We show further results in the appendix.
Fig.~\ref{fig:res_mv_materials} shows more examples of \shortTitle's output. The 21 frames are represented by 5 novel views sampled from an orbit around the object. The model is capable of generating a 3D consistent surface representation as evident in the normal maps and preservation of fine details, e.g. of the wheelchair. Illumination is successfully disentangled from the basecolor and roughness and metallic maps contain the spatial variance expected from the RGB views or ground truth given in Fig.~\ref{fig:res_single_materials}. The %
relighting results indicate that a physically plausible material prediction is achieved which, given the correct illumination, can reproduce the ground truth.
Compared to the other methods, the roughness and metallic from Stable Material (SM)~\cite{litmanMaterialFusionEnhancingInverse2024} are smoother than the ones from RGB$\leftrightarrow$X, but our results are overall closer to the ground truths. Most other methods use a monocular prior for the normal generation or are trained with annotations from a pre-trained model~\cite{kocsisIntrinsicImageDiffusion2023,zengRGBX2024}.
\input{figs/reb_app_25d_relighting_comparison}

\inlinesection{Runtime.}
\shortTitle~generates 21 views at 576$\times$576 in ~20s. While other methods may be faster per frame, they require minutes to generate full sequences. Our 3D reconstruction takes ~15 mins., with the ~3 min overhead vs. SV3D due to the added PBR optimization.

\inlinesection{Ablation study.} 
We ablate different aspects of~\shortTitle~in terms of reconstruction quality and present quantitative results computed on a subset of the Poly Haven~\cite{polyhavenPolyHaven} dataset in Tab.~\ref{tab:ablation_quantitative} in addition to the visual examples in Fig.~\ref{fig:issues}. Every ablated component contributes significantly to the final reconstruction quality.

\inlinesection{Limitations.}
Currently, our model focuses on object centric images, limiting its applicability to general video applications.
Furthermore, our PBR representation cannot represent more complex materials such as transparent objects.
Enhancing the material and illumination complexity in the diffusion denoising process pose interesting future work.

%% file: tables/res_single_material.tex
\begin{table}
\titlecaptionof{table}{Single frame material prediction}{Given a single RGB image the corresponding material parameters basecolor, roughness and metallic are generated and compared to GT from our Poly Haven test set. For \OURS we only evaluate the output for the condition frame. Results are averaged over 3 samples.}
\label{tab:single_material}
\resizebox{\linewidth}{!}{ %
\Huge
\begin{tabular}{@{}lcccccccc@{}}
\multirow{2}*{Method} & \multicolumn{4}{c}{Basecolor / Albedo} & \multicolumn{2}{c} {Roughness-Metallic} & \multicolumn{2}{c} {Normal}
\\\cmidrule(lr){2-5}\cmidrule(lr){6-7}\cmidrule(lr){8-9}
 & PSNR\textuparrow & SSIM\textuparrow & LPIPS\textdownarrow & FID\textdownarrow & PSNR\textuparrow  & RMSE\textdownarrow & PSNR\textuparrow & RMSE\textdownarrow
\\
\midrule
IID~\cite{kocsisIntrinsicImageDiffusion2023} & 8.62 & 0.66 & 0.49 & 120.17 & 11.80 & 0.33 & 12.48 & 0.28 \\
SM~\cite{litmanMaterialFusionEnhancingInverse2024} & \secondbest{20.59} & \secondbest{0.86} & \secondbest{0.072} & \secondbest{31.21}  & \secondbest{19.1} & \secondbest{0.16} & - & -\\
RGB$\leftrightarrow$X~\cite{zengRGBX2024} & 16.01 & 0.83 & 0.12 & 57.9 & 17.4 & \secondbest{0.16} & 6.96 & 0.45\\
\textbf{\OURS} (ours) & \best{28.68} & \best{0.92} & \best{0.037} & \best{18.3} & \best{25.36} & \best{0.09} & \best{27.57} & \best{0.05}\\
\end{tabular}
} %
\end{table}

%% file: figs/res_mv_comparison.tex
\begin{figure}[tb]
\centering
\includegraphics[width=\linewidth]{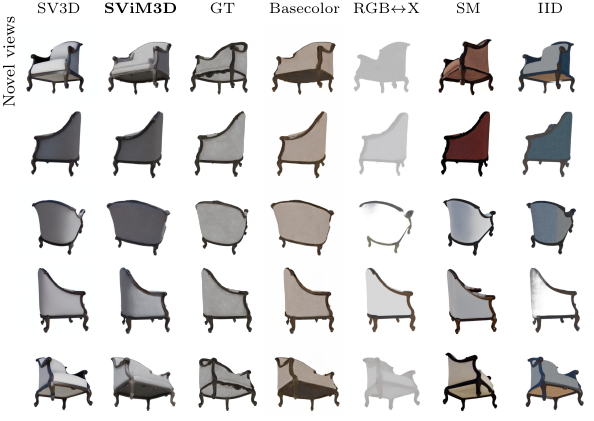}
\vspace{-.75em}
\titlecaption{Multi-view consistency}{We compare the generated materials from different neural diffusion priors in a multi-view setting. SV3D~\cite{voletiSV3DNovelMultiview2024} shows multi-view consistent RGB output similar to \shortTitle~that also generates multi-view consistent Basecolor. Generating albedo maps on top of the SV3D views using RGB$\leftrightarrow$X~\cite{zengRGBX2024}, StableMaterial (SM) of MaterialFusion~\cite{litmanMaterialFusionEnhancingInverse2024} or Intrinsic Image Diffusion (IID)~\cite{kocsisIntrinsicImageDiffusion2023} yields inconsistent results compared to the GT.}
\label{fig:res_mv_comparison}
\vspace{-.75em}
\end{figure}

%% file: figs/res_multiview_materials.tex
\begin{figure}[t]
\begin{center}
\includegraphics[trim={10.5cm 0cm 0cm 0cm},clip,width=\linewidth]{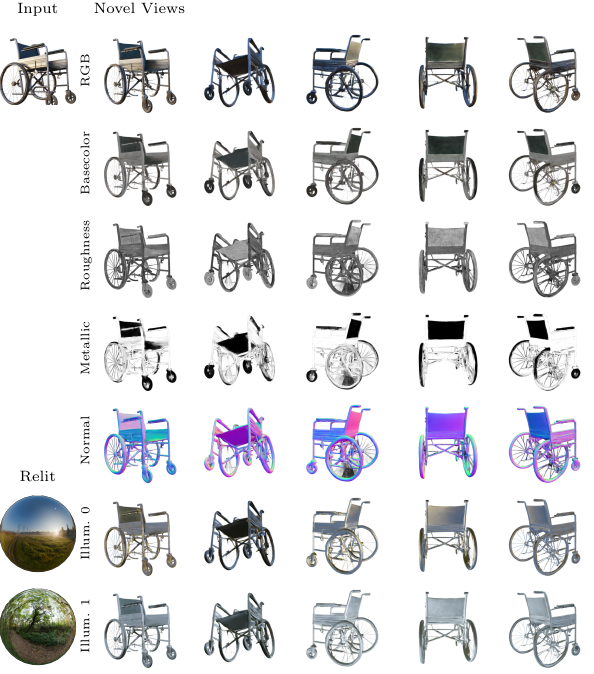}
\end{center}
\vspace{-1em}
\titlecaption{Multi-view PBR materials}{Given the input image \shortTitle~generates multi-view consistent novel views with corresponding basecolor, roughness, metallic and normal maps. These can directly be used to generate views under novel illumination. We show 5 samples from a generated orbit and two new illumination settings as examples. The objects are sourced from our Poly Haven~\cite{polyhavenPolyHaven} test dataset. Please find additional results in the supplementary material.
}
\label{fig:res_mv_materials}
\end{figure}

%% file: figs/res_single_material.tex
\begin{figure*}[ht]
\begin{center}
\includegraphics[width=\textwidth]{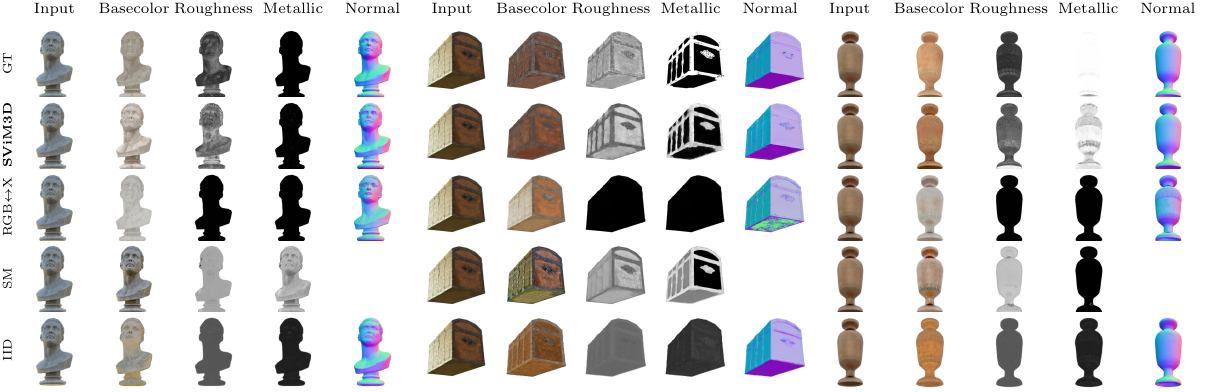}
\end{center}
\vspace{-1em}
\titlecaption{Single image PBR materials}{We compare the generated materials from different neural diffusion priors for a single image from the Poly Haven~\cite{polyhavenPolyHaven} test set. Besides the GT rendering and \shortTitle~(ours) results from RGB$\leftrightarrow$X~\cite{zengRGBX2024}, StableMaterial (SM) of MaterialFusion~\cite{litmanMaterialFusionEnhancingInverse2024} and Intrinsic Image Diffusion (IID)~\cite{kocsisIntrinsicImageDiffusion2023} are presented. Note that IID uses monocular normals that are separately generated and SM does not provide any normals.}
\label{fig:res_single_materials}
\end{figure*}

%% file: tables/res_mv_rgb_albedo.tex
\begin{table}
\titlecaptionof{table}{Multi-view NVS with material parameters}{Given a single RGB image a multi-view orbit around the scene is generated with corresponding PBR materials and normals. We compare RGB NVS and albedo generation as stand-in for PBR materials against rendered GT on our Poly Haven test set. Additionally, GT illumination is used to reproduce the RGB radiance from the predicted materials and normals for \OURS. Results are averaged over 3 samples for all 21 frames.}
\vspace{-.75em}
\label{tab:mv_rgb_albedo}
\resizebox{0.98\linewidth}{!}{ %
\Huge
\begin{tabular}{@{}lcccccccc@{}}
Method & PSNR\textuparrow & SSIM\textuparrow & LPIPS\textdownarrow & FID\textdownarrow & CLIPS\textuparrow  & CMMD\textdownarrow
\\
\midrule
RGB radiance 21 frames
\\\cmidrule(lr){1-1}
SV3D~\cite{voletiSV3DNovelMultiview2024} & 18.41 & 0.83 & 0.097 & \secondbest{7.8} & \secondbest{0.84} & \secondbest{1.06} \\
\textbf{\OURS} (ours) & \secondbest{19.57} & \secondbest{0.85} & \best{0.089} & \best{6.93} & \best{0.85} & 1.12\\
\textbf{\OURS} (ours) 2.5D relit & \best{19.99} & \best{0.87} & \best{0.089} & 15.15 & 0.83 & \best{0.08}\\
\toprule\toprule
Basecolor / Albedo 21 frames
\\\cmidrule(lr){1-1}
SV3D + IID~\cite{kocsisIntrinsicImageDiffusion2023} & 15.62 & 0.76 & 0.18 & 28.41 & \secondbest{0.81} & 1.81\\
SV3D + RGB$\leftrightarrow$X~\cite{zengRGBX2024} & 15.15 & \secondbest{0.83} & 0.11 & 22.13 & 0.80 & \secondbest{1.05}\\
SV3D + SM~\cite{litmanMaterialFusionEnhancingInverse2024} & \secondbest{18.12} & \secondbest{0.83} & \secondbest{0.10} & \secondbest{17.22} & \secondbest{0.81} & \best{0.96}\\
\textbf{\OURS} (ours) & \best{18.27} & \best{0.85} & \best{0.09} & \best{9.42} & \best{0.82} & 1.16\\
\end{tabular}
} %
\end{table}

%% file: tables/ablation_stanford_orb.tex
\begin{table}[t]
\begin{minipage}[t]{0.49\linewidth}
\titlecaptionof{table}{\footnotesize 3D reconstruction ablation}{We ablate different key aspects of our pipeline using a subset of the Poly Haven test set.}
\label{tab:ablation_quantitative}
\resizebox{\linewidth}{!}{ %
\begin{tabular}{lcc}
Configuration & PSNR\textuparrow & SSIM\textuparrow \\
\midrule
w/o homography correction & 13.7 & 0.76 \\
w/o environment lighting & 15.0 & 0.81 \\
w/o view masking & 17.8 & 0.84 \\
\OURS full & \best{22.4} & \best{0.90} \\
\end{tabular}
}
\end{minipage}
\hfill
\begin{minipage}[t]{0.48\linewidth}
\titlecaptionof{table}{\footnotesize Stanford Orb}{Novel view synthesis (NVS) and relighting evaluated on Stanford Orb~\cite{kuang2023stanfordorb}.}
\label{tab:stanford_orb}
\resizebox{\linewidth}{!}{ %
\begin{tabular}{@{}ccccc@{}}
Method & \multicolumn{2}{c} {NVS} & \multicolumn{2}{c} {Relighting}
\\
 & PSNR\textuparrow & SSIM\textuparrow & PSNR\textuparrow & SSIM\textuparrow
\\
\midrule
SF3D~\cite{boss_sf3d2024} & 16.81 & 0.74 & 20.1 & 0.88 \\
\shortTitle & \best{19.34} & \best{0.80} & \best{21.86} & \best{0.90}
\end{tabular}
}
\end{minipage}
\end{table}

%% file: figs/reb_app_25d_relighting_comparison.tex
\begin{figure}[tb]
\centering
\includegraphics[width=\linewidth]{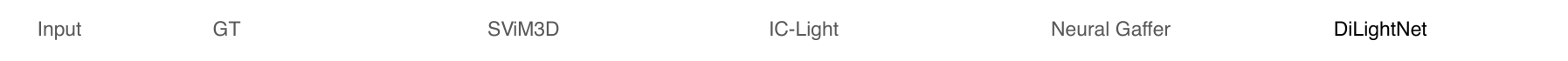}\\
\includegraphics[width=\linewidth]{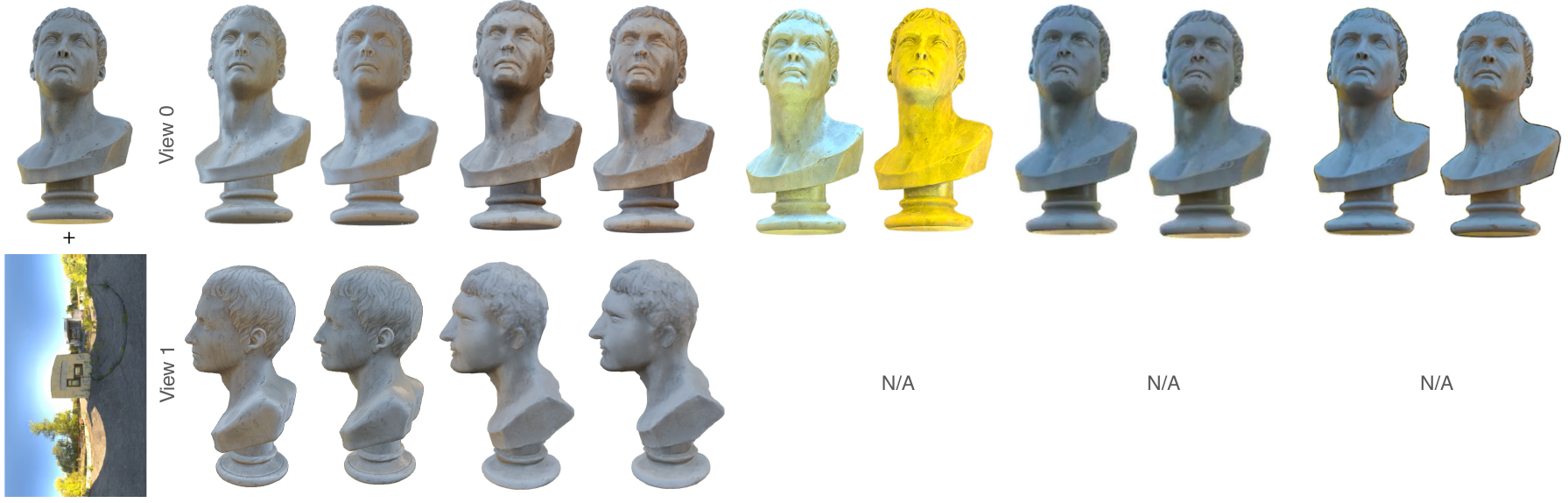}
\vspace{-1.0em}
\titlecaption{Relighting comparison}{We compare image-based relighting of recent diffusion-based methods IC-Light~\cite{zhang2025scaling_iclight}, Neural-Gaffer~\cite{jin_neural_2024} and DiLightNet~\cite{zeng2024dilightnet} against~\shortTitle~and the synthetic ground truth (GT) on examples from Poly Haven~\cite{polyhavenPolyHaven} data.}
\label{fig:app_25d_relighting_comparisons}
\vspace{-.75em}
\end{figure}

%% file: 10_conclusion.tex
\vspace{-2mm}
\section{Conclusion}
\label{sec:conclusion}
\vspace{-2mm}

We present \shortTitle, the first foundational multi-view material model. Given a conditioning image and user-defined camera path, \shortTitle~jointly predicts multi-view consistent RGB colors, spatially varying PBR material parameters and surface normals. We demonstrate the quality and consistency of \shortTitle's outputs by employing them as pseudo ground truth in a 3D reconstruction pipeline, showing that it enables high-quality 3D reconstructions in this ill-posed setting. 
We adapt a video diffusion model, introducing key modifications to the network architecture and training data to enable simultaneous material prediction. We also introduce several innovations in 3D reconstruction to correct multi-view inconsistencies, and add fast differentiable environment-based lighting.
Our extensive experiments demonstrate the state-of-the-art performance of \shortTitle~in several tasks related to novel view and material synthesis. We hope \shortTitle~can serve as a foundational model for future research on multi-view consistent material generation.

\ifreview
\else
\section*{Acknowledgements}
\vspace{-2mm}
This work has been partially funded by the Deutsche Forschungsgemeinschaft (DFG, German Research Foundation) under Germany’s Excellence Strategy – EXC number 2064/1 – Project number 390727645 and SFB 1233, TP 02  -  Project number 276693517.
\fi

%% file: 12_appendix.tex
\section*{Overview}
\ifarxiv
In the supplement to ...
\fi
\ifcamera
In the supplement to \longerTitle~(\shortTitle), a foundational multi-view material model with camera control, we first expand the background on video diffusion models and neural fields, add information on the optimization and finally present more results, in-depth analysis and applications of our method. 
Please also consider watching the \textbf{supplemental video} that gives an overview of this work and contains further visual results.
\fi

\section{Additional Background}

\subsection{Video Diffusion Denoising}
The conditioning image is concatenated to the noisy latent state input $z_t$ at noise timestep $t$. The CLIP-embedding~\cite{pmlr-v139-radford21a} matrix of the conditioning image is provided to the cross-attention layers of each transformer block as its key and value. The camera poses, represented as angles $e_i$ and $a_i$ as well as the noise timestep $t$ are encoded into sinusoidal position embeddings. The camera pose embeddings are linearly transformed and added to the noise timestep embedding. The result is added to each residual block's output features after being run through another linear layer to match the feature dimension as in SV3D~\cite{voleti2024sv3d}. 
\input{figs/app_multi_samples}

\subsection{Coordinate-based MLPs and NeRF}~\cite{mildenhall2020}
NeRFs~\cite{mildenhall2020} 
use a dense neural network to model a continuous function that takes 3D location $\vect{x} \in \mathbb{R}^{3}$ and view direction $\vect{d} \in \mathbb{R}^3$ and outputs a view-dependent output color $\vect{c} \in \mathbb{R}^{3}$ and volume density $\sigma \in \mathbb{R}$.
A camera ray $r(t) = \vect{o} + t \vect{d}$ is cast into the volume, with ray origin $\vect{o} \in \mathbb{R}^3$ and view direction $\vect{d}$. 
The final color is then approximated via numerical quadrature of the integral: $\vect{\hat{c}}(\vect{r})=\int_{t_n}^{t_f} T(t)\sigma(t)\vect{c}(t)\, dt$ with $T(t) = \exp (-\int_{t_n}^{t} \sigma(t)\, dt )$, using the near and far bounds of the ray $t_n$ and $t_f$ respectively~\cite{mildenhall2020}.

\section{Optimization}

\subsection{UNet training details}
We pre-compute latents and CLIP-embeddings~\cite{pmlr-v139-radford21a} for all training data.
The RGB color rendering is composed on a solid random color or white, the basecolor AOV stays always on white. The other outputs keep their black backgrounds. We follow the EDM framework and use the diffusion loss for fine-tuning described by Blattmann \etal~\cite{blattmannStableVideoDiffusion2023}.
We employ Flash Attention v2~\cite{dao2022flashattention,daoFlashAttention2FasterAttention2023} to keep the memory footprint low such that a batch size of two is still possible for 21 frames on similar hardware to the SV3D~\cite{voleti2024sv3d} training.

\inlinesection{Guidance.}
Compared to a conventional video generation with a reference frame as the starting point we have circular orbits both starting and ending close to the reference view. To reduce over-sharpening caused by classifier-free-guidance (CFG)~\cite{ho2021classifierfree} we also adapt a triangular CFG scaling similar to the one proposed in~\cite{voletiSV3DNovelMultiview2024} where the guidance scale is adapted based on the distance to the reference view. 

\subsection{Geometry regularization.}
We adopt several geometric priors to regularize the reconstructed shape. Firstly we supervise the normal using the predicted normal maps. Especially during the beginning of the NeRF optimization this supervision loss is strictly enforced eliminating the need for any additional monocular prior.
Since our normal maps generally contain more detail than can be represented by the mesh representation, starting from the second half of phase 1, we additionally optimize a bump map represented by a small auxiliary field conditioned on the coordinate embeddings from DMTet. 
A bilateral smoothness loss is also added to the normals in phase 1 and increased during phase 2. Similarly, we utilize the smooth depth loss from RegNeRF~\cite{Niemeyer2021Regnerf}.
While the supervision loss with the pseudo-GT (pGT) and the photometric rendering loss are high in the beginning of the NeRF reconstruction (Phase 1) we slowly increase the weight of the LPIPS~\cite{zhang2018lpips} over the course of the reconstruction ultimately dominating the reconstruction at the end of Phase 1. Our homography correction scheme is also added in Phase 1 after an initial warmup phase of 400 steps.
In Phase 2 the LPIPS loss is slowly reduced a little and bilateral smoothness regularizers increased in weight to clean up remaining noise.

\subsection{View dependent masking}
We normalize the masks by the maximum value over all views and apply a smoothstep function $f_s$ followed by a gamma correction to smoothly clip to the range of 0 to 1 and to steer the mask contrast.

\subsection{Homography correction}
To make the optimization more robust to outlier views where the image is warped wrongly due to homogeneous image regions or complex edge features, we introduce a masking scheme in Phase 2. Based on the loss difference in the albedo map, it is decided if the current view is warped or not. If a view is consistently masked, then $H_i$ is reinitialized and further refined.

\section{Further results}
In the following section we provide additional results including evaluation on additional datasets and qualitative comparisons related to the reconstruction pipeline.
\input{figs/app_additional_mv_materials}

\input{figs/app_sf3d_comparison}

\input{tables/suppl_table_mv}
\input{figs/res_mv_comparisons_2}

\subsection{Overview of baseline methods}
Intrinsic Image Diffusion (IID)~\cite{kocsisIntrinsicImageDiffusion2023} is one of the first works to explore diffusion models for PBR material estimation. Their model outputs albedo, roughness and metallic parameters for a single frame. 
Originally trained on interior scenes, it has also been applied to general 3D reconstruction~\cite{duGSIDIlluminationDecomposition2024}.
MaterialFusion~\cite{litmanMaterialFusionEnhancingInverse2024} proposes a 2D material denoising diffusion prior based on StableDiffusion 2.1~\cite{rombachHighResolutionImageSynthesis2022} with the same output as above but trained on object centric data. They employ an SDS based optimization to achieve 3D asset generation.
Finally, RGB$\leftrightarrow$X~\cite{zengRGBX2024} released a latent image diffusion model that can generate PBR data as part of their material- and lighting-aware neural rendering pipeline. Their 
material model can generate either albedo, roughness, metallic or diffuse irradiance maps conditioned on a single image and a text prompt to select the task.
Significantly faster is SF3D~\cite{boss_sf3d2024} which is based on a transformer decoder architecture like LRM~\cite{hong2023lrm,wei2024meshlrm}.
Since the 3D reconstruction code for SV3D~\cite{voleti2024sv3d} is not publicly available at the time of writing we decide to compare against SF3D instead.
As evident in Fig~\ref{fig:app_sf3d_comparison} SF3D's material model is limited as it does not allow for spatially-varying roughness and metallic values. This poses a severe limitation for real-world objects composed from multiple materials. Our spatially-varying parametrization yields shading results closer to the GT. Tab.~\ref{tab:baselines} gives a high-level overview of the features available in the compared methods. \shortTitle~is the only one offering RGB view synthesis and material synthesis as a multi-view task with joint spatially-varying PBR and normal prediction as well as 3D reconstruction of a textured mesh.

\input{tables/baselines}
\input{tables/baselines_2d}

\subsection{Additional multi-view material results}
In Fig~\ref{fig:app_mv_materials}, Fig.~\ref{fig:res_mv_materials2} and Fig~\ref{fig:app_mv_materials_gso} we show additional raw outputs of our diffusion model given reference images from multiple datasets. \shortTitle~generates plausible material maps for a variety of object classes and surface materials. The high metallic value in Fig~\ref{fig:app_mv_materials_gso} is questionable in a physical sense but apparently helps the model to represent the specific shine of the dinosaur figure which might correspond to the way an artist might work in this case.
In Fig.~\ref{fig:app_mv_materials_gt} we compare the generated material maps to the ground truth AOVs from synthetic data. Despite the ambiguity the model is able to predict plausible solutions also reflected in the RMSE values in Tab.~\ref{tab:single_material}.
In addition to our newly introduced Poly Haven~\cite{polyhavenPolyHaven} object dataset we also evaluate our model on a test split of the recently introduced BlenderVault dataset~\cite{litmanMaterialFusionEnhancingInverse2024} in Tab.~\ref{tab:mv_eval_blendervault}. The results are consistent with our evaluation on Poly Haven verifying the plausabiliy of our test results.

\subsection{Quantitative evaluation across views}
Fig.~\ref{fig:mv_error} compares the mean error across all generated views between all evaluated models from Tav.~\ref{tab:mv_rgb_albedo}. Our method consistently yields the best results over all views, although it varies depending on the camera view. The observation that the side views are the most challenging generations might be explained by the occurrence of more extreme angle configurations in the context of the surface shading. Traditionally, grazing angles and samples close to object boundaries can lead to inconsistencies in 3D reconstruction~\cite{zollhoferStateArt3D2018} and generation might suffer from similar effects. Additionally, Tab.~\ref{tab:met3r} shows the results of MeT3R~\cite{asim24met3r}, a view consistency metric based on the recently introduced DUST3R~\cite{WangDUSt3R_2024_CVPR} for calibration free 3D point cloud reconstruction. The metric also reflects the improved multi-view consistency in~\shortTitle~compared to the SV3D~\cite{voleti2024sv3d} baselines.

\subsection{3D reconstruction}
\begin{figure}[htb]
  \centering
\includegraphics[width=0.7\linewidth]{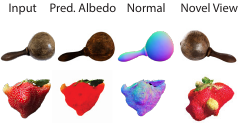}
\vspace{-.75em}
\titlecaption{Real-world results}{Example generations from casual smartphone captures of a shaker instrument and a strawberry.}
\label{fig:reb_real_world}
\end{figure}

Fig.~\ref{fig:app_reconstruction_example} illustrates our single image to 3D reconstruction pipeline using an example image from our test set. Starting with the multi-view novel view synthesis with material parameters and surface normals, the output is lifted to a 3D representation, first a NeRF~\cite{mildenhall2020}, then a polygon mesh. It is worth noting that the material parameters are well preserved thanks to our pseudo GT supervision. Finally, the mesh can be rendered under novel illumination, again.
We show additional 3D reconstruction results in Fig.~\ref{fig:reb_additional_reconstruction_results}. Fig.~\ref{fig:reb_real_world} features two generations conditioned on a smartphone capture illustrating in-the-wild performance.

\input{figs/reb_additional_reconstructions}

\input{figs/app_25d_relighting}
\subsection{Multiple samples}
Fig.~\ref{fig:app_multi_samples} compares three samples of denoising process given the same condition image. It is visible that there is some diversity in the predictions while they still all represent physically plausible solutions in the context of the conditioning given the underconstrained task. The diversity of the deviations increases the further the camera moves away from the condition frame, of course. 

\subsection{3D Geometry}
We evaluate the quality of the reconstructed geometry using Chamfer distance and Intersection over Union (IoU) against ground truth point clouds provided by the Google Scanned Objects (GSO)~\cite{gso2022} dataset and report the results in Tab.~\ref{tab:3d_eval}. We select a random subset of 80 real-world objects for the comparison against SF3D~\cite{boss_sf3d2024}. Compared to the feed-forward architecture of SF3D can our reconstruction method fail in rare cases where some views do not align for some reason. This is reflected in the slightly lower scores. In cases where reconstruction succeeds the quality is visually very close, often keeping a bit finer detail in the case of \shortTitle~at the expensive of some additional noise (see also Fig~\ref{fig:app_sf3d_comparison}).

\input{figs/app_reconstruction_example}
\input{figs/app_mv_materials_gso}
\input{figs/app_mv_generated}

\inlinesection{RGB only view synthesis}
Using the SV3D~\cite{voleti2024sv3d} baseline without PBR material prediction yields lower quality results also for the RGB color generation as reported in Tab.~\ref{tab:mv_rgb_albedo}. We argue that enforcing reasoning over illumination as part of the material estimation also helps the generation of consistent lighting in the RGB views.

\input{tables/3d_eval}

\input{tables/app_mv_blendervault}

\input{figs/material_editing_roughness}
\subsection{Relighting}
In Fig.~\ref{fig:app_25d_relighting} we give additional insights into our 2.5D relighting approach. We show a metallic and plastic surface lit by different rotations of the spherical environment map. Using all the generated material channels and the normal directions we can achieve dynamic direct illumination at real-time speed. We also present the intermediate illumination representation used in our deferred shading pipeline. Our pipeline also enables material editing as further analyzed in Fig.~\ref{fig:app_material_editing}. Fig~\ref{fig:relighting_examples} shows examples for different illumination directions and camera views.
To achieve indirect illumination, a full 3D reconstruction can be completed.

\input{figs/relighting_comparison}
\input{figs/relighting_examples}
\inlinesection{Relighting comparison}
We present additional results from our 2.5D relighting pipeline in Fig.~\ref{fig:relighting_comparison}.
As baselines we use IC-Light~\cite{zhang2025scaling_iclight}, Neural Gaffer~\cite{jin_neural_2024} and DiLightNet~\cite{zeng2024dilightnet}, three diffusion based methods for image-based relighting recently introduced. In Tab.~\ref{tab:baselines_relighting} we give an overview of the feature sets of all relighting methods.
Neural Gaffer supports environment map inputs as conditioning which is fed as low and high dynamic range representation. IC-Light provides image editing based on a background image. And DiLightNet adds radiance hints to the conditioning via environment maps. In our comparison we preprocess the environment maps to serve the methods, respectively.
We compare the results against the GT obtained from our 2.5D rendering pipeline here, using the synthetic PBR material maps. \shortTitle~is the only model capable of joint novel view synthesis and relighting. This is reflected in better multi-view consistency and fewer artifacts like the residual highlight in the example of Neural-Gaffer. IC-Light generally generated high-contrast output which is difficult to edit in real-world use cases.
\begin{figure}[htb]
\centering
    \centering
    \includegraphics[width=0.6\linewidth]{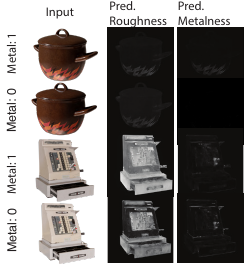}
  
    \vspace{-.75em}
    \titlecaption{Glossiness vs. Metalness ambiguity}{Examples from our generated test cases and the corresponding model predictions.}
      
    \label{fig:reb_material_ambiguity}
    \end{figure}

\inlinesection{3D relighting application}
As shown in Fig.~\ref{fig:app_reconstruction_example} as well as Fig.~\ref{fig:teaser} the 3D reconstructed models can be easily integrated into new environments thanks to the PBR materials. Using a path tracer global illumination effects can then be achieved, too. Please find additional dynamic relighting and scene integration examples in the supplemental video.

\input{figs/app_mv_material_gt}

\inlinesection{Analysis of ambiguous materials}
We constructed a small dataset of pathological test cases for the ambiguity between metallic and glossy plastic surfaces. 
In over 90\% of the cases a low roughness value with near zero metalness is predicted. The predictions of higher values often are for objects that would usually have metal in their material.
See Fig.~\ref{fig:reb_material_ambiguity} for a visual example.
These findings can be explained by dataset bias.

%% file: figs/app_multi_samples.tex
\begin{figure}[tb]
\centering
\includegraphics[width=\linewidth]{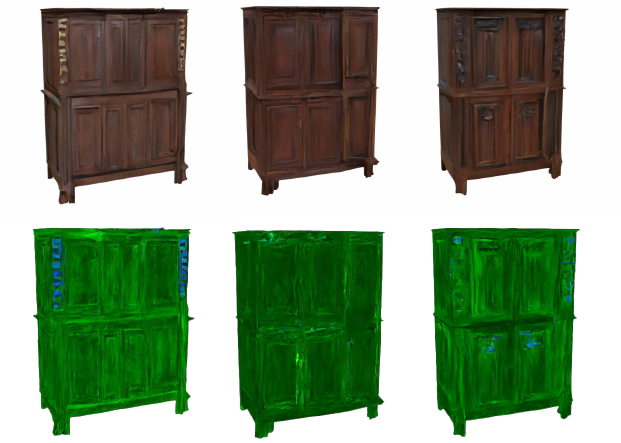}
\vspace{-.75em}
\titlecaption{Multiple samples}{Demonstrating the stochastic sampling process by taking three samples with the same condition image. For views that are less constrained by the conditioning diverse examples can be generated depending on the initial noise. Note, that the roughness and metallic parameters (blue and green here) are consistent with the RGB predictions, though.}
\label{fig:app_multi_samples}
\vspace{-.75em}
\end{figure}

%% file: figs/app_additional_mv_materials.tex
\begin{figure*}[t]
\centering
\includegraphics[width=\textwidth]{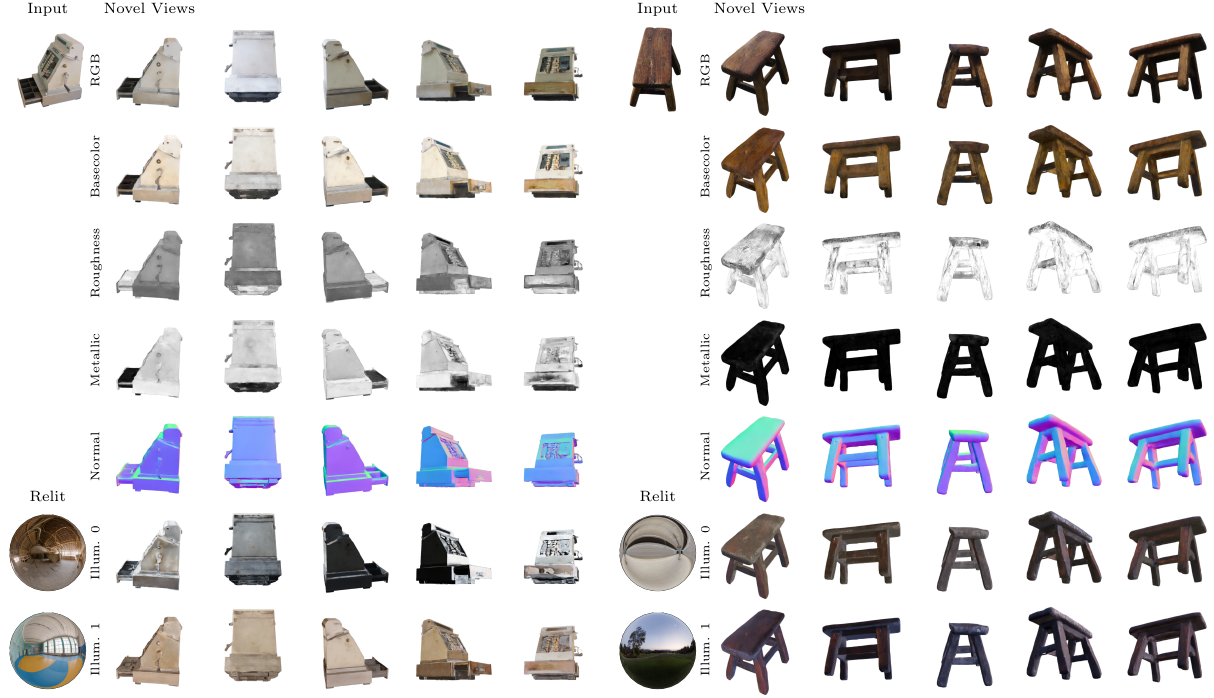}
\vspace{-.75em}
\titlecaption{Multi-view material prediction}{Additional examples from the Poly Haven~\cite{polyhavenPolyHaven} test dataset. \shortTitle~successfully converts a single image to a sequence of novel views with spatially-varying PBR material parameters and surface normals. These can directly be used to relight the novel views as shown in the two bottom rows.}
\label{fig:app_mv_materials}
\vspace{-.75em}
\end{figure*}

%% file: figs/app_sf3d_comparison.tex
\begin{figure}
\begin{minipage}{\linewidth}
    \centering
     \begin{subfigure}[b]{0.32\textwidth}
         \centering
         \includegraphics[trim=100 100 100 200,clip,width=\textwidth]{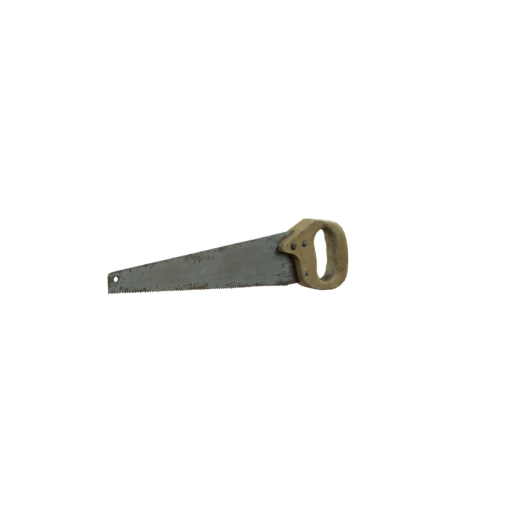}
     \end{subfigure}
     \hfill
     \begin{subfigure}[b]{0.32\textwidth}
         \centering
         \includegraphics[trim=50 50 80 200,clip,width=\textwidth]{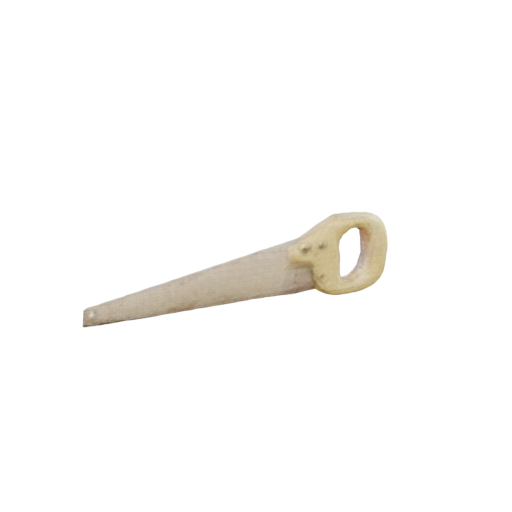}
     \end{subfigure}
     \hfill
     \begin{subfigure}[b]{0.32\textwidth}
         \centering
         \includegraphics[trim=50 100 50 200,clip,width=\textwidth]{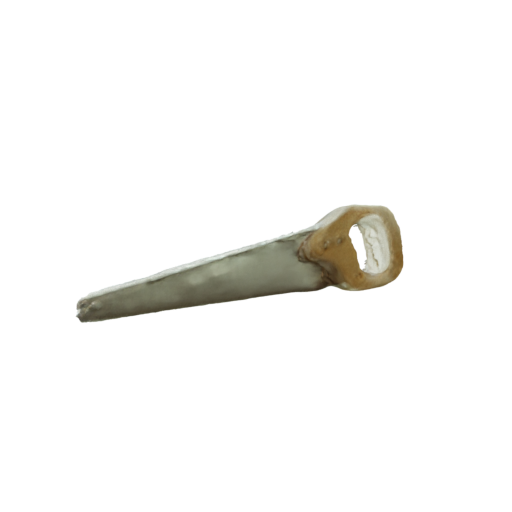}
     \end{subfigure}
     \\
     \begin{subfigure}[b]{0.32\textwidth}
         \centering
         \includegraphics[trim=100 100 100 200,clip,width=\textwidth]{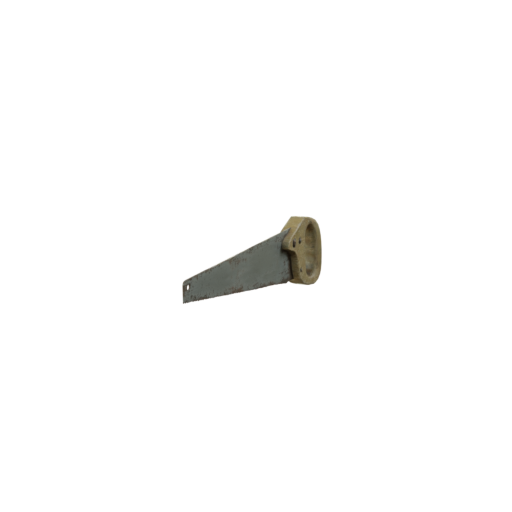}
          \caption{\footnotesize GT rendering}
     \end{subfigure}
     \hfill
     \begin{subfigure}[b]{0.32\textwidth}
         \centering
         \includegraphics[trim=90 60 90 200,clip,width=\textwidth]{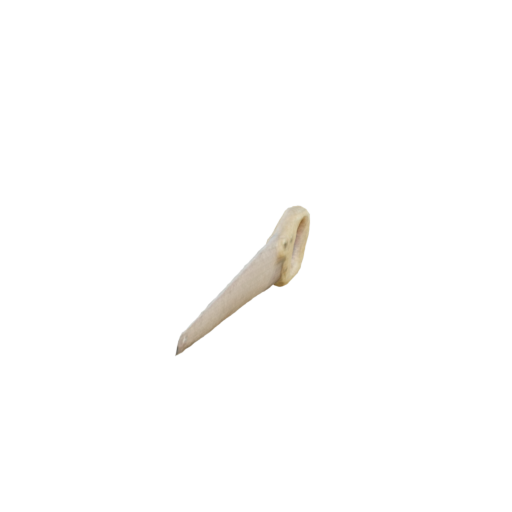}
         \caption{\footnotesize SF3D}
     \end{subfigure}
     \hfill
     \begin{subfigure}[b]{0.32\textwidth}
         \centering
         \includegraphics[trim=80 60 80 200,clip,width=\textwidth]{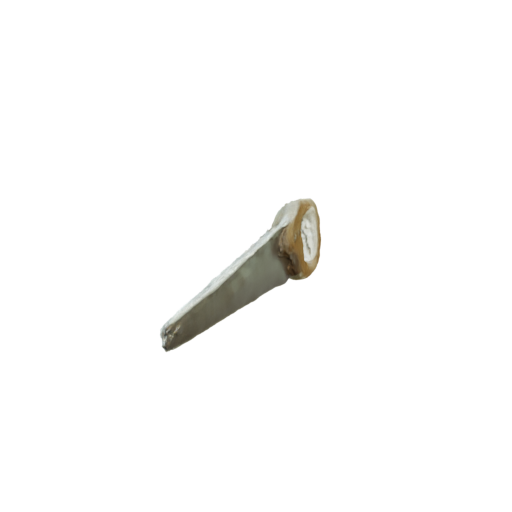}
         \caption{\footnotesize \shortTitle}
     \end{subfigure}
        \titlecaption{Material parametrization}{Compared to SF3D~\cite{boss_sf3d2024}, a recent method for single image to 3D generation, our material model is able to replicate spatially-varying roughness and metallic parameters which help to represent real-world objects realistically.}
        \label{fig:app_sf3d_comparison}
\end{minipage}
\end{figure}

%% file: tables/suppl_table_mv.tex
\begin{figure}[ht]
\centering %
\begin{minipage}{0.48\linewidth} %
\titlecaptionof{table}{\footnotesize View consistency}{Multi-view consistency evaluated using MEt3R~\cite{asim24met3r} on the Poly Haven test data.}
\label{tab:met3r}
\resizebox{\linewidth}{!}{%
\begin{tabular}{lc}
\toprule
Method & MEt3R score\textuparrow \\
\midrule
SV3D RGB$\leftrightarrow$X~\cite{zengRGBX2024} & 0.54 \\
SV3D + IID~\cite{kocsisIntrinsicImageDiffusion2023} & 0.51 \\
SV3D + SM~\cite{litmanMaterialFusionEnhancingInverse2024} & 0.54 \\
\OURS & 0.57 \\
\bottomrule
\end{tabular}
}
\end{minipage}%
\hfill
\begin{minipage}{0.48\linewidth}

         \includegraphics[trim=0 2 0 0,clip,width=\textwidth]{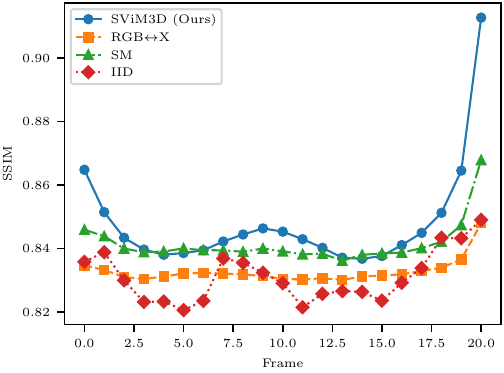}
\caption{\footnotesize \textbf{Multi-view error distribution.} We compare the SSIM results of the Basecolor prediction across frames over the Poly Haven test set.}
\label{fig:mv_error}
\end{minipage}
\end{figure}
\hspace{-4pt}

%% file: figs/res_mv_comparisons_2.tex
\begin{figure}[t]
\begin{center}
\includegraphics[trim={0cm 0cm 10.5cm 0cm},clip,width=\linewidth]{images/results/multi_view_materials_results_v07.pdf}
\end{center}
\vspace{-1em}
\titlecaption{Multi-view PBR materials}{Given the input image \shortTitle~generates multi-view consistent novel views with corresponding basecolor, roughness, metallic and normal maps. These can directly be used to generate views under novel illumination. We show 5 samples from a generated orbit and two new illumination settings as examples. The objects are sourced from our Poly Haven~\cite{polyhavenPolyHaven} test dataset. Please find additional results in the supplementary material.
}
\label{fig:res_mv_materials2}
\end{figure}

%% file: tables/baselines.tex
\begin{table}
\titlecaptionof{table}{Baseline Methods}{Features of existing methods used in our evaluation compared to \shortTitle.}
\label{tab:baselines}
\resizebox{\linewidth}{!}{ %
\Huge
\begin{tabular}{@{}lcccccc@{}}
Method & RGB NVS & Multi-view & Joint PBR & Spatially-varying PBR & Normals & Textured mesh
\\
\midrule
SV3D~\cite{voletiSV3DNovelMultiview2024} & \cmark & \cmark & \xmark & \xmark & \xmark & \cmark \\
SF3D~\cite{boss_sf3d2024} & \xmark & \xmark & \cmark & \xmark & \cmark & \cmark\\
IID~\cite{kocsisIntrinsicImageDiffusion2023} & \xmark & \xmark & \cmark & \cmark & \xmark & \xmark \\
RGB$\leftrightarrow$X~\cite{zengRGBX2024} & \xmark & \xmark & \xmark & \cmark & \cmark & \xmark \\
SM~\cite{litmanMaterialFusionEnhancingInverse2024} & \xmark & \xmark & \cmark & \cmark & \xmark & \cmark \\
\shortTitle & \cmark & \cmark & \cmark & \cmark & \cmark & \cmark \\
\end{tabular}
}
\end{table}

%% file: tables/baselines_2d.tex
\begin{table}
\titlecaptionof{table}{Baseline Methods Relighting}{Features of existing methods for image based relighting compared to \shortTitle.}
\label{tab:baselines_relighting}
\resizebox{\linewidth}{!}{ %
\Huge
\begin{tabular}{@{}lccccccc@{}}
Method & LDR output & HDR output & Global Illum & NVS & Multi-view & Material Editing & Interactive speed
\\
\midrule
IC Light~\cite{zhang2025scaling_iclight} & \cmark & \xmark & \cmark & \xmark & \xmark & \xmark & \xmark \\
Neural Gaffer~\cite{jin2024neural_gaffer} & \cmark & \xmark & \cmark  & \xmark & \xmark & \xmark & \xmark\\
\shortTitle & \cmark & \cmark & \xmark & \cmark & \cmark & \cmark & \cmark \\
\end{tabular}
}
\end{table}

%% file: figs/reb_additional_reconstructions.tex
\begin{figure}[tb]
\centering
\includegraphics[trim={0 3cm 0 0},clip,width=0.9\linewidth]{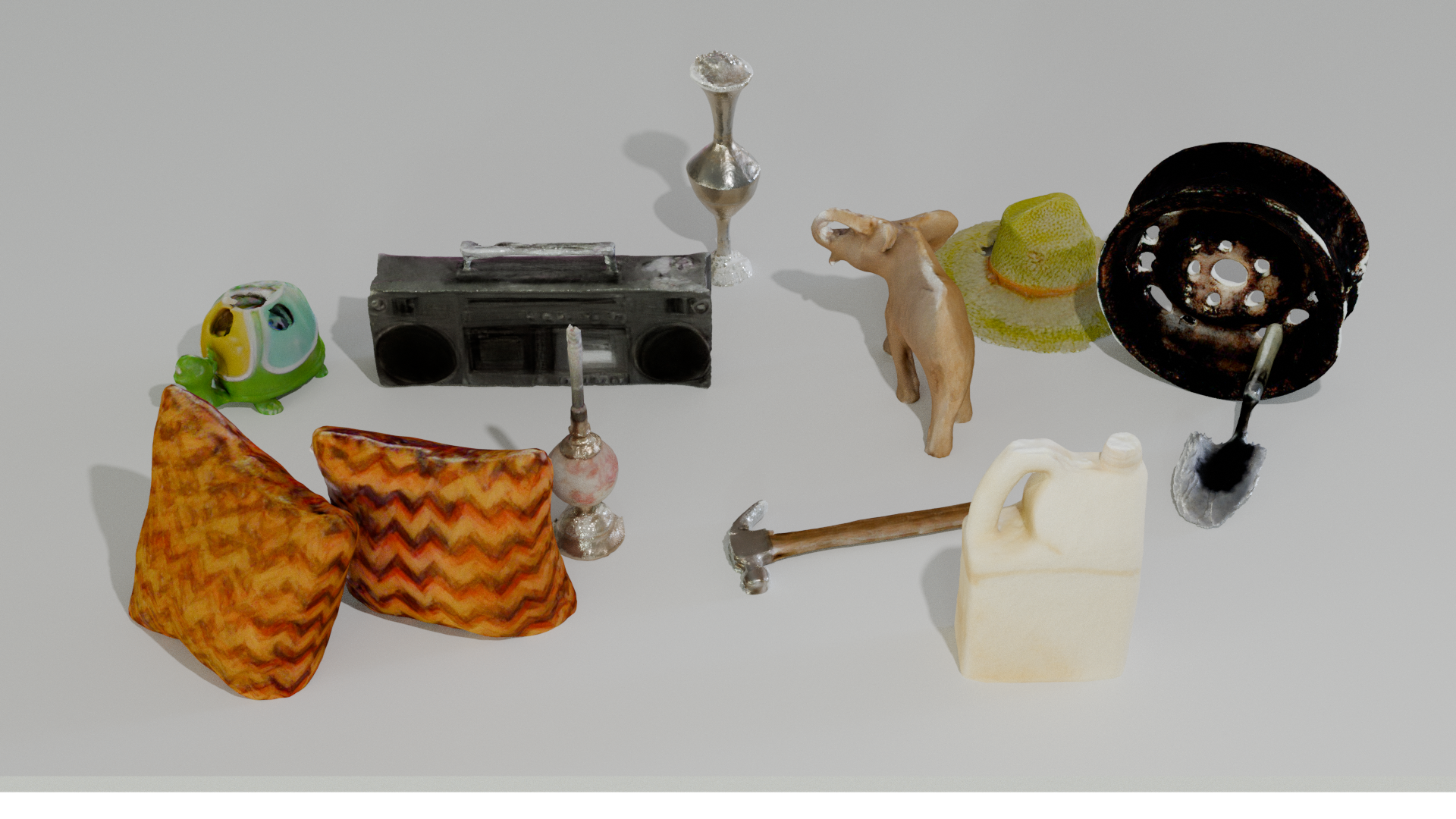}
\vspace{-.5em}
\titlecaption{More 3D reconstruction results}{Objects sourced from Poly Haven~\cite{polyhavenPolyHaven} and GSO~\cite{gso2022}, rendered in Blender.}
\label{fig:reb_additional_reconstruction_results}
\vspace{-.75em}
\end{figure}

%% file: figs/app_25d_relighting.tex
\begin{figure}[tb]
\centering
\includegraphics[width=\linewidth]{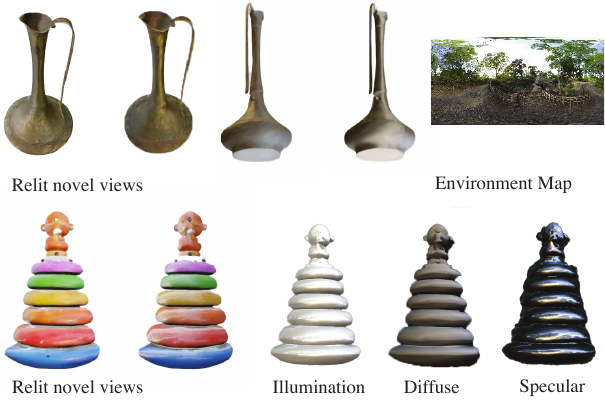}
\vspace{-.75em}
\titlecaption{2.5D Relighting}{Using the output of \shortTitle~and an environment map we can directly relight an object. We can use the same illumination representation and deferred shading as in the differentiable rendering pipeline.}
\label{fig:app_25d_relighting}
\vspace{-.75em}
\end{figure}

%% file: figs/app_reconstruction_example.tex
\begin{figure*}[t]
\begin{center}
\includegraphics[width=\textwidth]{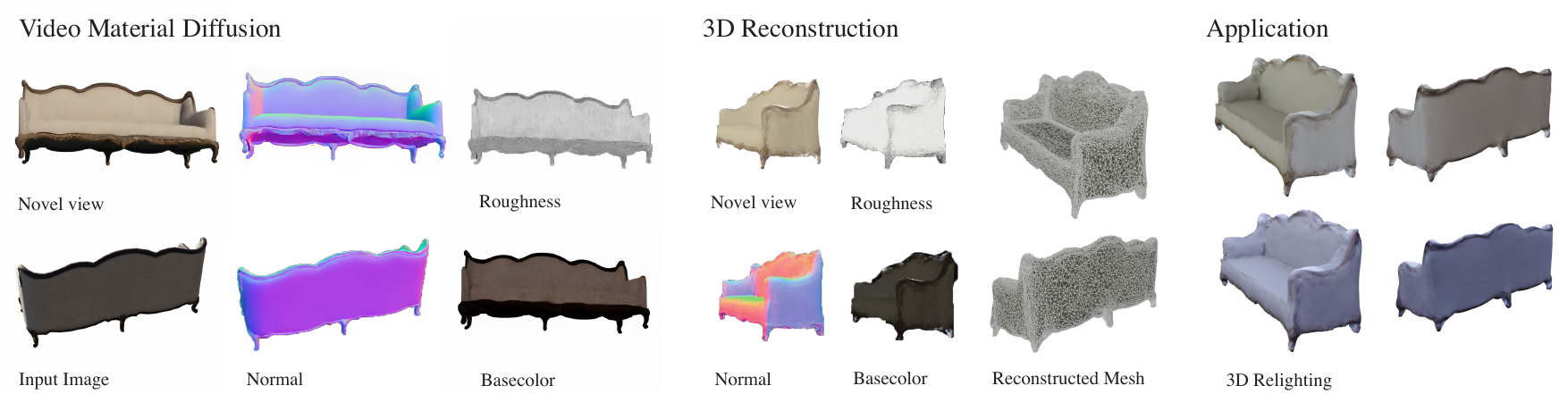}
\end{center}
\vspace{-1em}
\titlecaption{3D reconstruction example}{\shortTitle~'s pipeline starts with a single image at the bottom left. First novel views and the corresponding material parameters and surface normals are generated. Following, an intermediate 3D representation is optimized given the multi-view material prior. Finally, a 3D mesh can be extracted and integrated into downstream applications. Here we show an example from our Poly Haven~\cite{polyhavenPolyHaven} test dataset.}
\label{fig:app_reconstruction_example}
\end{figure*}

%% file: figs/app_mv_materials_gso.tex
\begin{figure*}[tb]
\centering
\includegraphics[width=\textwidth]{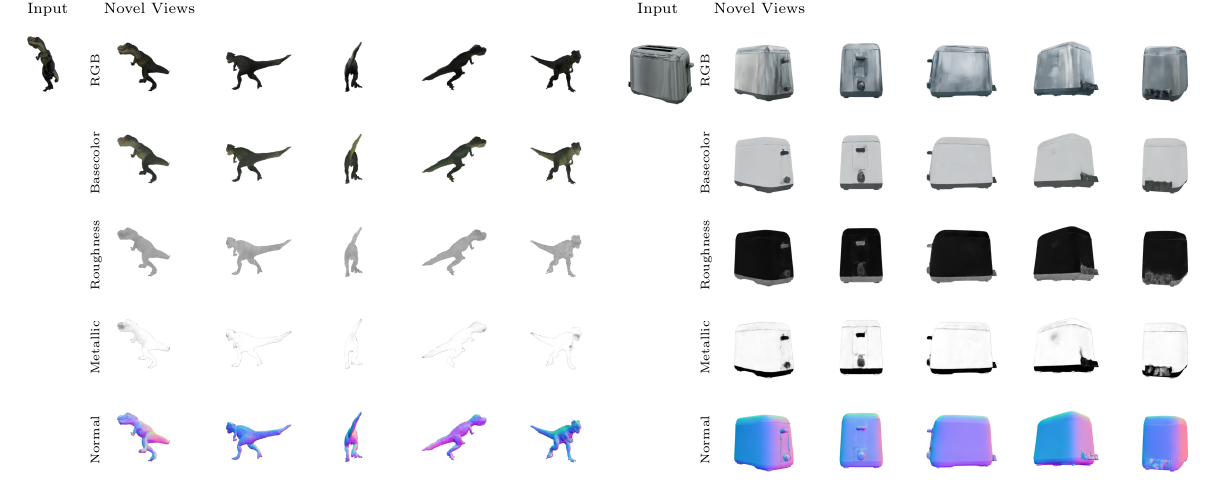}
\vspace{-.75em}
\titlecaption{Multi-view material examples from GSO}{Two objects from the GSO~\cite{gso2022} dataset representing common real-world houshold items. \shortTitle~generalizes well to this domain as long as the scene is object centric.}
\label{fig:app_mv_materials_gso}
\vspace{-.75em}
\end{figure*}

%% file: figs/app_mv_generated.tex
\begin{figure*}[tb]
\centering
\includegraphics[width=\textwidth]{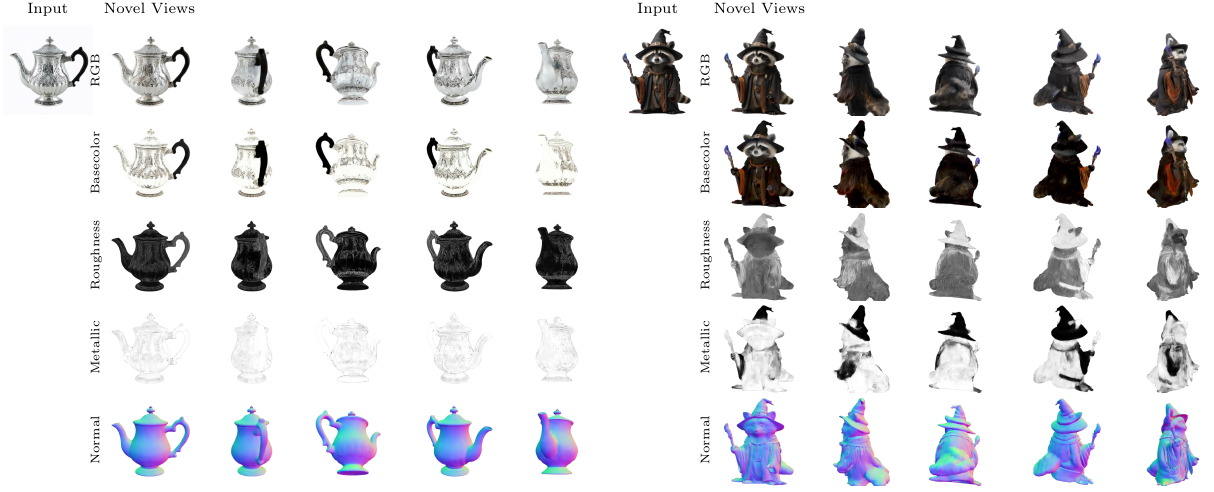}
\vspace{-.75em}
\titlecaption{Multi-view material examples from generated images.}{Multi-view generations conditioned on generated images from text-to-image models, a wizard raccoon and a silver teapot. \shortTitle~is capable of estimating plausible and view consistent results. The wizard raccoon is an out-of-distribution example due to the lack of stylized character models in the training data.}
\label{fig:app_mv_materials_generated}
\vspace{-.75em}
\end{figure*}

%% file: tables/3d_eval.tex
\begin{table}
\titlecaptionof{table}{3D reconstruction}{We evaluate the model against SF3D~\cite{boss_sf3d2024} on a subset of the Google Scanned Objects (GSO)~\cite{gso2022} featuring real-world household items. The mesh quality is reported as Chamfer distance and IoU compared to the scanned GT pointclouds.}
\label{tab:3d_eval}
\resizebox{0.7\linewidth}{!}{ %
\begin{tabular}{@{}lccc@{}}
\multirow{2}*{Method} & \multicolumn{2}{c} {3D Geometry}
\\\cmidrule(lr){2-3}
 & Chamfer\textdownarrow & IoU\textuparrow
\\
\midrule
SF3D~\cite{boss_sf3d2024} & 0.031 & 0.52 &  \\
\OURS & 0.034 & 0.48 &  \\
\end{tabular}
}
\end{table}

%% file: tables/app_mv_blendervault.tex
\begin{table}
\titlecaptionof{table}{Multi-view NVS with material parameters on BlenderVault dataset}{Given a single RGB image a multi-view orbit around the scene center is generated with corresponding PBR materials and normals. We compare RGB NVS and albedo / basecolor generation as stand-in for PBR materials against rendered GT on a subset (100 objects) of the BlenderVault dataset~\cite{litmanMaterialFusionEnhancingInverse2024}. We also compare against the MaterialFusion~\cite{litmanMaterialFusionEnhancingInverse2024} baseline on their single view prediction task.}
\vspace{-.75em}
\label{tab:mv_eval_blendervault}
\resizebox{0.98\linewidth}{!}{ %
\Huge
\begin{tabular}{@{}lcccccccc@{}}
Method & PSNR\textuparrow & SSIM\textuparrow & LPIPS\textdownarrow & FID\textdownarrow & CLIPS\textuparrow  & CMMD\textdownarrow
\\
\midrule
RGB radiance 21 images
\\\cmidrule(lr){1-1}
\textbf{\OURS} (ours) & 20.22 & 0.86 & 0.081 & 24.95 & 0.86 & 1.14\\
\toprule\toprule
Basecolor / Albedo 21 images
\\\cmidrule(lr){1-1}
\textbf{\OURS} (ours) & 19.80 & 0.86 & 0.08 & 40.0 & 0.81 & 1.08\\
\midrule
Basecolor single image (ref view)
\\\cmidrule(lr){1-1}
SM~\cite{litmanMaterialFusionEnhancingInverse2024} (\textit{from paper}) & 24.70 & 0.91 & - & - & - & -\\
\textbf{\OURS} (ours) & 27.35 & 0.92 & 0.05 & 46.0 & 0.83 & 1.08\\
\end{tabular}
} %
\end{table}

%% file: figs/material_editing_roughness.tex
\begin{figure}[tb]
\centering
\includegraphics[width=\linewidth]{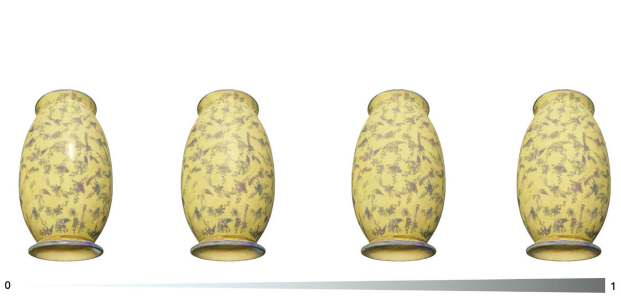}
\vspace{-.75em}
\titlecaption{Material editing}{The explicit material parameters of \shortTitle's output can be edited in a physically-plausible way and the result visualized using our rendering framework. In this example the material roughness is varied between almost zero and close to one while the original value is close to the version second to left.}
\label{fig:app_material_editing}
\vspace{-.75em}
\end{figure}

%% file: figs/relighting_comparison.tex
\begin{figure*}[tb]
\centering
\includegraphics[width=\linewidth]{images/applications/relighting_comparison_03_rebuttal_03.pdf}
\includegraphics[width=\linewidth]{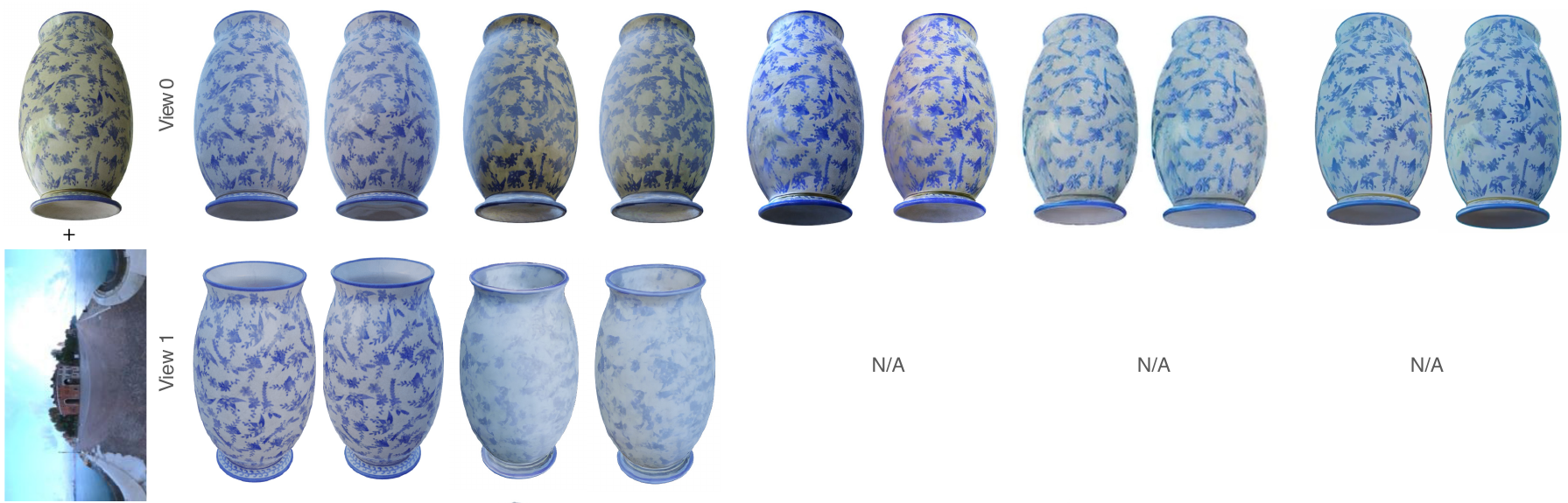}
\vspace{-.75em}
\titlecaption{Relighting comparison}{We compare image-based relighting results on an example object from the Poly Haven~\cite{polyhavenPolyHaven} dataset between the synthetic ground truth (GT), IC-Light~\cite{zhang2025scaling_iclight}, Neural-Gaffer~\cite{jin_neural_2024}, DiLightNet~\cite{zeng2024dilightnet} and \shortTitle~(ours).}
\label{fig:relighting_comparison}
\vspace{-.75em}
\end{figure*}

%% file: figs/relighting_examples.tex
\begin{figure}[tb]
\centering
\includegraphics[width=\linewidth]{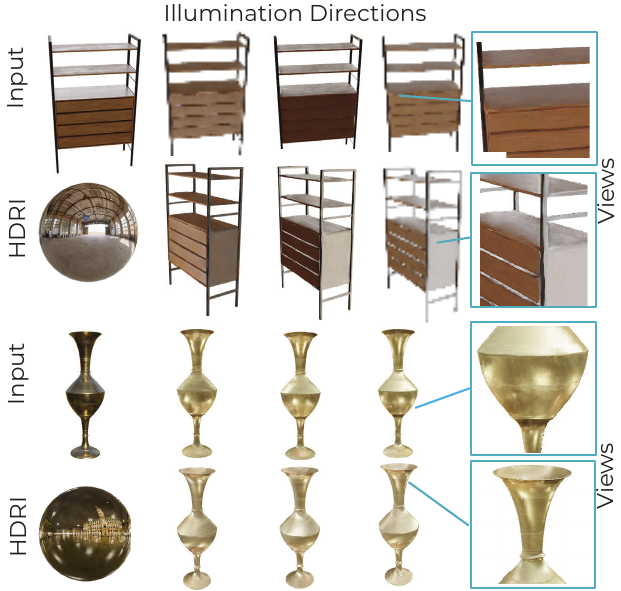}
\vspace{-.75em}
\titlecaption{Relighting}{Using the output of \shortTitle~and an environment map (HDRI) we can directly relight any view on the camera trajectory using our 2.5D approach.}
\label{fig:relighting_examples}
\vspace{-.75em}
\end{figure}

%% file: figs/app_mv_material_gt.tex
\begin{figure*}[tb]
\centering
 \resizebox{0.8\textwidth}{!}{
\includegraphics[width=0.8\textwidth]{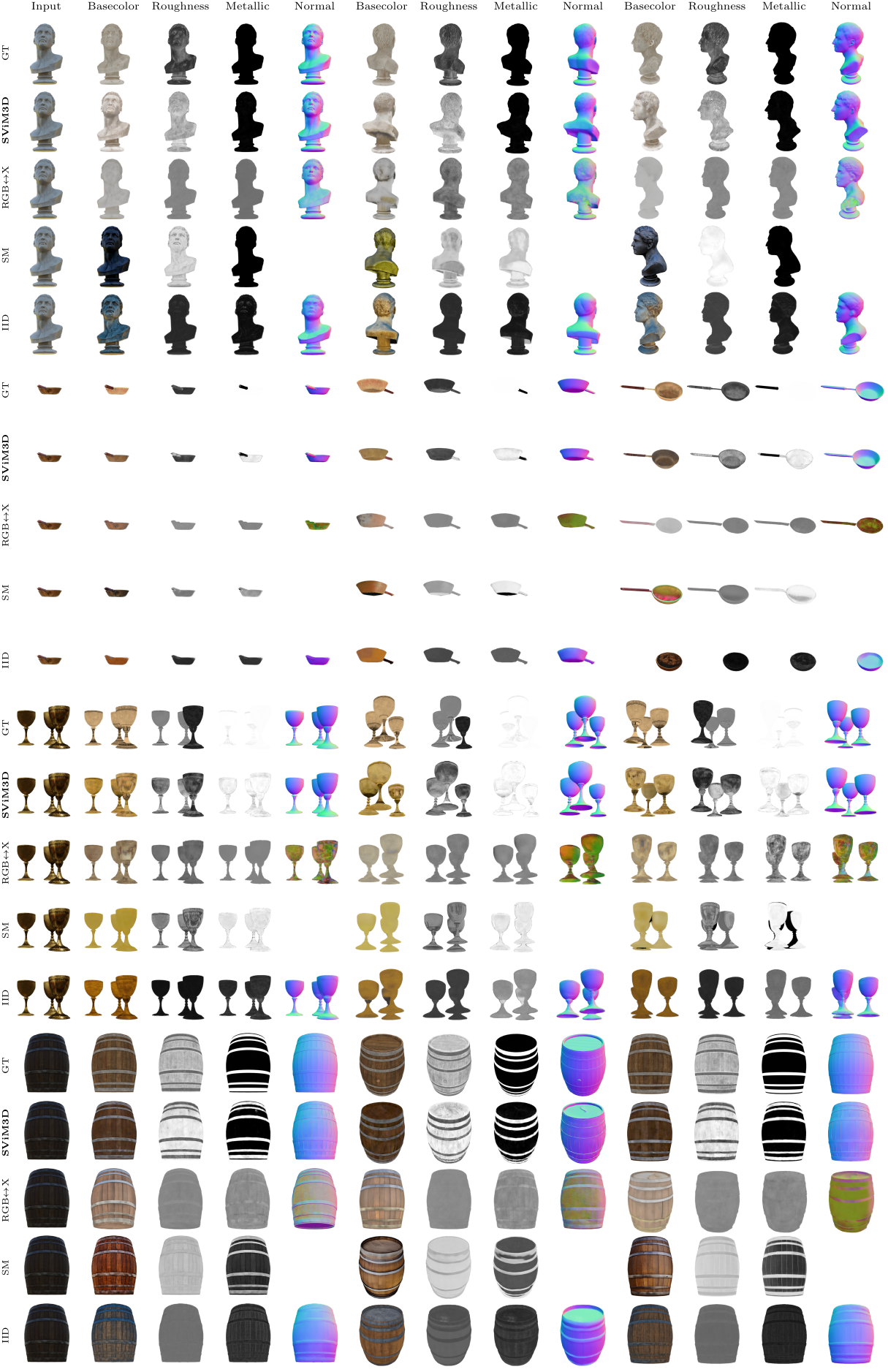}
}
\vspace{-.75em}
\titlecaption{Comparison of multi-view material generation on Poly Haven objects}{We compare generated materials of RGB$\leftrightarrow$X~\cite{zengRGBX2024}, StableMaterial (SM) of MaterialFusion~\cite{litmanMaterialFusionEnhancingInverse2024} and Intrinsic Image Diffusion (IID)~\cite{kocsisIntrinsicImageDiffusion2023} based on SV3D~\cite{voleti2024sv3d} generations and \shortTitle~for three views around the object against GT renders.}
\label{fig:app_mv_materials_gt}
\vspace{-.75em}

\end{figure*}